\documentclass[11pt,fleqn,twoside]{article}
\usepackage{amsfonts,amssymb,latexsym}
\makeatletter
\newcommand{\prava}[1]{\small\it
\begin{flushleft}
Copyright \copyright \ 1999 by  #1
\end{flushleft}}

\newcommand{\name}[1]{\begin{flushleft}
                       \LARGE \bf #1
                       \end{flushleft}\vspace{-3mm}}

\newcommand{\Author}[1]{\begin{flushleft}
                       \it #1 \end{flushleft}}

\newcommand{\Adress}[1]{\begin{flushleft}
                       \it #1 \end{flushleft}}

\newcommand{\Date}[1]{\begin{flushleft}
                      \small  \it #1 \end{flushleft}}

\newcommand{\ehkol}{Author \ name}
\newcommand{\ohkol}{Article \ name}
\renewcommand{\@evenhead}{
\hspace*{-3pt}\raisebox{-15pt}[\headheight][0pt]{\vbox{\hbox to \textwidth 
{\thepage \hfil \ehkol}\vskip4pt \hrule}}}
\renewcommand{\@oddhead}{
\hspace*{-3pt}\raisebox{-15pt}[\headheight][0pt]{\vbox{\hbox to \textwidth 
{\ohkol \hfil \thepage}\vskip4pt\hrule}}}
\renewcommand{\@evenfoot}{}
\renewcommand{\@oddfoot}{}

\newcommand{\be}{\begin{equation}}
\newcommand{\ee}{\end{equation}}
\newcommand{\ba}{\hspace*{-5pt}\begin{array}}
\newcommand{\ea}{\end{array}}

\newcommand{\ds}{\displaystyle}
\makeatother

\newcommand{\ga}{{\bf a}}
\newcommand{\g}{\mbox{\boldmath$g$}}
\newcommand{\mbL}{\mbox{\boldmath$L$}}
\newcommand{\mbM}{\mbox{\boldmath$M$}}
\newcommand{\rR}{{\rm R}}
\newcommand{\rA}{{\rm A}}
\newcommand{\rH}{{\rm H}}
\newcommand{\rP}{{\rm P}}
\newcommand{\rS}{{\rm S}}
\newcommand{\rW}{{\rm W}}
\newcommand{\bA}{{\bf A}}
\newcommand{\bB}{{\bf B}}
\newcommand{\bg}{{\bf g}}
\newcommand{\bM}{{\bf M}}
\newcommand{\bR}{{\bf R}}
\newcommand{\bS}{{\bf S}}
\newcommand{\bT}{{\bf T}}
\newcommand{\bL}{\mbox{\boldmath$L$}}
\newcommand{\cA}{{\mathcal A}}
\newcommand{\cB}{{\mathcal B}}
\newcommand{\cC}{{\mathcal C}}
\newcommand{\cE}{{\mathcal E}}
\newcommand{\cF}{{\mathcal F}}
\newcommand{\cL}{{\mathcal L}}
\newcommand{\cR}{{\mathcal R}}
\newcommand{\cT}{{\mathcal T}}
\newcommand{\cX}{{\mathcal X}}
\newcommand{\qed}{\rule{3mm}{3mm}}

\newtheorem{theorem}{Theorem}[section]
\newtheorem{proposition}[theorem]{Proposition}
\newtheorem{lemma}[theorem]{Lemma}
\newtheorem{definition}[theorem]{Definition}

\begin{document}
\thispagestyle{empty}
\setcounter{page}{411}
\renewcommand{\ehkol}{Yu.B. Suris}
\renewcommand{\ohkol}{$r$-Matrices for Relativistic Deformations of Integrable 
Systems}

\begin{flushleft}
\footnotesize \sf Journal of Nonlinear Mathematical Physics \qquad
1999, V.6, N~4, \pageref{suris-fp}--\pageref{suris-lp}.
\hfill {\sc Article}
\end{flushleft}

\vspace{-5mm}

\renewcommand{\footnoterule}{}
{\renewcommand{\thefootnote}{}
 \footnote{\prava{Yu.B. Suris}}}

\name{{\mathversion{bold}$r$}-Matrices for Relativistic Deformations \\
of Integrable  Systems}\label{suris-fp}

\Author{Yuri B. SURIS}

\Adress{Fachbereich Mathematik, Sekr. MA 8-5,
Technische Universit\"at Berlin, \\
Str. des 17. Juni 136, 10623 Berlin, Germany\\
E-mail:  suris@sfb288.math.tu-berlin.de}

\Date{Received June 06, 1999; Revised July 05, 1999; Accepted August 05, 1999}

\begin{abstract}
\noindent
We include the relativistic lattice KP hierarchy,
introduced by Gibbons and Kupershmidt, into the $r$-matrix framework.
An $r$-matrix account of the nonrelativistic lattice KP hierarchy 
is also provided for the reader's convenience. All relativistic constructions
are regular one-parameter perturbations of the nonrelativistic ones.
We derive in a simple way the linear Hamiltonian structure of the relativistic
lattice KP, and f\/ind for the f\/irst time its quadratic Hamiltonian structure.
Amasingly, the latter turns out to coincide with its nonrelativistic 
counterpart (a phenomenon, known previously only for the simplest case of
the relativistic Toda lattice).
\end{abstract}

\renewcommand{\theequation}{\thesection.\arabic{equation}}
\setcounter{equation}{0}
\section{Introduction}\label{Introduction}

One of the basic objects in the theory of integrable lattices (or 
dif\/ferential-dif\/ference systems) is the so called {\it lattice KP hierarchy}
\cite{suris:K1,suris:K3}. This is a system of an inf\/inite number of commuting f\/lows
on the phase space consisting of an inf\/inite number of f\/ields
\begin{equation}\label{GTL introd phase sp}
b=(b_k)_{k\in {\mathbb Z}}, \qquad 
a^{(j)}=\left(a_k^{(j)}\right)_{k\in {\mathbb Z}} \quad (j\ge 1).
\end{equation} 
The simplest (``f\/irst'') f\/low of this hierarchy is governed by the following
equations of motion:
\begin{equation}\label{GTL introd}
\left\{\begin{array}{l}
\dot{b}_k=a_k^{(1)}-a_{k-1}^{(1)}, 
\vspace{2mm}\\
\dot{a}_k^{(j)}=a_k^{(j)}(b_{k+j}-b_k)+
\left(a_k^{(j+1)}-a_{k-1}^{(j+1)}\right). 
\end{array}\right.
\end{equation}
A proper language for such inf\/inite-dimensional systems is the 
dif\/ferential-dif\/ference calculus developed in \cite{suris:K1,suris:K3}.
However, all essential properties of this hierarchy hold also for its
f\/inite-dimensional reductions described as follows.

First of all, we shall consider here only systems with a f\/inite number
of f\/ields. We~obtain an $m$-f\/ield system, if we set $a_k^{(j)}=0$ for all 
$j\ge m$. We shall use the notation ${\rm TL}_m$ for this reduced system,
because it is a direct generalization of the {\it Toda lattice}, which appears
in this framework as the f\/irst nontrivial case ${\rm TL}={\rm TL}_2$ with two
f\/ields. 

Further, we restrict ourselves here to the case of periodic or open-end
boundary conditions, when all f\/ields contain only a f\/inite number $N$ of
variables, so that in (\ref{GTL introd phase sp}) one should replace 
$k\in{\mathbb Z}$ through $1\le k\le N$. More precisely, in case of periodic
boundary conditions all subscripts in (\ref{GTL introd}) are supposed to
belong to ${\mathbb Z}/N{\mathbb Z}$, and in case of open-end boundary conditions
we enforce
\[
\ba{l}
b_k=0\quad {\rm for}\quad  k<1 \quad  {\rm and}\quad  {\rm for}\quad k>N,
\vspace{2mm}\\
a_k^{(j)}=0\quad {\rm for}\quad  k<1 \quad {\rm and}\quad  {\rm for}\quad
k>N-j.
\ea
\]
One of the outstanding features of this hierarchy is integrability; another
one is its bi-Hamiltonian structure. What we want to address in the present
paper, is an $r$-matrix theory as the (probably most direct) route towards
understanding these both properties.

Actually, the $r$-matrix interpretation of the lattice KP hierarchy is a
more or less established thing nowadays 
\cite{suris:AM,suris:RSTS,suris:S1,suris:O}. What is really 
new in this paper, is a similar interpretation of the {\it relativistic lattice
KP hierarchy}. This appeared in \cite{suris:OR,suris:GK}
as a natural generalization of
the relativistic Toda lattice, invented by Ruijsenaars~\cite{suris:R}. The relativistic
ansatz of \cite{suris:GK} (referred to later on as ``the f\/irst construction'') leads 
to a one-parameter perturbation of the lattice KP hierarchy. 
An interesting ``splitting'' phenomenon takes place: to each 
(polynomial) f\/low of the lattice KP hierarchy there correspond two f\/lows
of its relativistic conterpart, one of them being still polynomial, and another
one rational in coordinates. It turns out that the rational relativistic
perturbation of the ``f\/irst'' f\/low~(\ref{GTL introd}) remains nice and 
elegant:
\begin{equation}\label{GRTL GK introd}
\left\{\begin{array}{l}
\ds \dot{b}_k=\frac{a_k^{(1)}}{1+\alpha b_{k+1}}-
\frac{a_{k-1}^{(1)}}{1+\alpha b_{k-1}},
\vspace{3mm}\\
\ds \dot{a}_k^{(j)}=a_k^{(j)}\left(\frac{b_{k+j}}{1+\alpha b_{k+j}}
-\frac{b_k}{1+\alpha b_k}\right)+
\left(\frac{a_k^{(j+1)}}{1+\alpha b_{k+j+1}}-
\frac{a_{k-1}^{(j+1)}}{1+\alpha b_{k-1}}\right),
\end{array}\right.
\end{equation}
while the polynomial relativistic perturbation is described by a system of
messy equations, right-hand side of each one depending on all f\/ields.
In the $m$-f\/ield situation we shall denote the f\/low (\ref{GRTL GK introd})
and the whole hierarchy attached to it by ${\rm RTL}_m^{(-)}(\alpha)$. It is a
direct generalization of the ``minus f\/irst'' f\/low of the relativistic Toda
hierarchy.

It is known that this relativistic lattice KP hierarchy consists of 
commuting f\/lows, and also one Hamiltonian structure was known for it 
previously \cite{suris:GK,suris:K2,suris:K3}. In the
present work we f\/ind also the second compatible Hamiltonian structure,
and, moreover, give an $r$-matrix account for both. Needless to say,
that this automatically implies also the integrability of this hierarchy.
Amasingly, while the f\/irst Hamiltonian structure of the relativistic
hierarchy is a perturbation of its nonrelativistic counterpart, the second
Hamiltonian structures of the both hierarchies literally coincide!

The relativistic ansatz of \cite{suris:GK} leads to Lax equations of a somewhat
non-standard type. We introduce a certain (invertible) gauge transformation,
such that the transformed hierarchy possesses the standard Lax representations.
In other words, it may be considered as included into the lattice KP 
hierarchy. It is a highly non-trivial situation, if one takes into account
that the lattice KP may be seen also as a limiting case of the relativistic
lattice KP! This gauge transformed relativistic hierarchy (referred to as
``the second construction'') also contains two perturbations, a polynomial
one and a rational one, for each nonrelativistic f\/low, but this time
the polynomial f\/lows are simpler. For example, the polynomial ``f\/irst'' f\/low
in the second relativistic construction reads:
\begin{equation}\label{GRTL introd}
\left\{\begin{array}{l}
\ds \dot{b}_k=(1+\alpha b_k)\left(a_k^{(1)}-a_{k-1}^{(1)}\right), 
\vspace{2mm}\\
\ds \dot{a}_k^{(j)}=a_k^{(j)}\left(b_{k+j}-b_k+
\alpha a_{k+j}^{(1)}-\alpha a_{k-1}^{(1)}\right)+
\left(a_k^{(j+1)}-a_{k-1}^{(j+1)}\right).
\end{array}\right.
\end{equation}
We denote this f\/low, as well as the whole corresponding hierarchy, in the
$m$-f\/ield situation as ${\rm RTL}_m^{(+)}(\alpha)$; it serves as a direct 
generalization of the ``f\/irst'' f\/low of the relativistic Toda hierarchy.

Remarkably, in the case $m=2$, i.e. of the relativistic Toda, both 
constructions lead to one and the same hierarchy, but this is no more the
case for $m>2$!

Also the second relativistic lattice KP turns out to be bi-Hamiltonian
and integrable; we give in this paper an $r$-matrix interpretation of one
of its Hamiltonian structures.

The paper is built as follows. In Sect.~\ref{Sect linear r-matrix 
brackets}--\ref{Sect quadr br on big algebra} we recall several notions
and results from the $r$-matrix theory (those of Sect.~\ref{Sect gen lin 
r-matrix} being to a certain extent new). Sect.~\ref{Sect notations} is
devoted to f\/ixing the basic notations. The lattice KP hierarchy and its
linear and quadratic invariant Poisson structures are dealt with in
Sect.~\ref{Sect GTL Lax}--\ref{Sect GTL quadratic r-matrix}. The similar
discussion of the f\/irst relativistic construction is contained in
Sect.~\ref{Sect alternative GRTL}--\ref{Sect quadratic bracket for GRTL GK} 
(so that the presentation of really new results starts in Sect.~\ref{Sect 
r-matrix for GRTL of GK}). Further, we introduce the second relativistic
construction in Sect.~\ref{Sect GRTLs relation}, \ref{Sect GRTL}, and 
discuss its Hamiltonian properties in Sect.~\ref{Sect qudratic bracket for 
GRTL}. As illustrations, we provide concrete results for the three-f\/ield
systems in Sect.~\ref{Sect GTL example}, \ref{Sect GRTL example}. A f\/inal
discussion takes place in Sect.~\ref{Sect conclusions}.

\setcounter{equation}{0}
\section{Linear {\mathversion{bold}$r$}-matrix Poisson structure}
\label{Sect linear r-matrix brackets}
We f\/irst recall the (by now classical) construction of the linear $r$-matrix
structure on the dual to a Lie algebra. Let $\g$ be a Lie algebra, carrying 
a nondegenerate scalar product $\langle\cdot,\cdot\rangle$, allowing to 
identify the dual space $\g^*$ with $\g$. In what follows we shall not
distinguish between $\g$ and $\g^*$, but try to use the letters $X$,  $Y$, $Z$ 
consequently for the elements of $\g$, and the letters $L$, $M$ for the elements
of $\g^*$. Suppose that the scalar product is invariant with respect to the Lie 
bracket in $\g$, i.e.
\[
\langle X,[Y,Z]\rangle = \langle [X,Y],Z\rangle\qquad
\forall \; X,Y,Z\in\g.
\] 
Recall that the gradient $\nabla\varphi:\g\mapsto\g$ of a smooth 
function $\varphi$ on $\g$ is def\/ined via
\[
\langle\nabla\varphi(L),M\rangle=\frac{d}{d\varepsilon}\varphi(L+
\varepsilon M)\Bigr|_{\varepsilon=0}\qquad \forall \; M\in\g.
\]
Recall also that Ad-invariant functions $\varphi$ on $\g$ are characterized
by the property
\begin{equation}\label{Ad inv}
[\nabla\varphi(L),L]=0\qquad \forall \; L\in\g.
\end{equation}
(Of course, we mean here actually ${\rm Ad}^*$-invariant functions on $\g^*$,
characterized by 
\[
{\rm ad}^* \nabla\varphi(L)\cdot L=0;
\] 
we shall consequently avoid such remarks from now on).

\begin{definition} {\bf \cite{suris:STS}}
 Let $\rR$ be a linear operator on $\g$.
A {\bfseries \itshape linear {\mathversion{bold}$r$}-matrix bracket} on $\g$
corresponding to the operator 
$\rR$ is defined by:
\begin{equation}\label{lin br}
\{\varphi,\psi\}_1(L)=\frac{1}{2}\langle
[\rR(\nabla \varphi(L)),\nabla \psi(L)]+[\nabla \varphi(L),\rR(\nabla \psi(L))],L\rangle.
\end{equation}
If this is indeed a Poisson bracket, it will be denoted by ${\rm PB}_1(\rR)$. 
\end{definition}
\begin{theorem} {\bf \cite{suris:STS}} A sufficient condition for (\ref{lin br})
to define a Poisson bracket is given by the {\bfseries \itshape modified Yang--Baxter 
equation} for the operator $\rR$:
\begin{equation}\label{mYB}
[\rR(X),\rR(Y)]-\rR\left([\rR(X),Y]+[X,\rR(Y)]\right)=-\alpha [X,Y]
\qquad\forall X,Y\in\g,
\end{equation}
where $\alpha$ is an arbitrary constant. This equation is denoted 
${\rm mYB}(\rR;\alpha)$.
\end{theorem}
Actually, (\ref{mYB}) is a suf\/f\/icient condition for 
\begin{equation}\label{R algebra}
[X,Y]_{\rR}=\frac{1}{2}\left([\rR(X),Y]+[X,\rR(Y)]\right)
\end{equation}
to def\/ine a new Lie bracket on $\g$, and then (\ref{lin br}) is nothing 
but a Lie--Poisson bracket on~$\g^*$ corresponding to this new Lie algebra 
structure on $\g$.

One of the most important properties of the linear $r$-matrix bracket is the 
following~one.

\begin{theorem}\label{lin bracket Ham equations} {\bf \cite{suris:STS}}
a) Hamiltonian equations of motion on $\g$ 
corresponding to an ${\rm Ad}$-invariant Hamilton function $\varphi$, 
have the Lax form
\begin{equation}\label{Lax with R lin}
\dot{L}=[L,C],\qquad C=\frac{1}{2}\rR(\nabla\varphi(L)).
\end{equation}

b) ${\rm Ad}$-invariant functions on $\g$ are in involution with
respect to the bracket ${\rm PB}_1(\rR)$. 
\end{theorem}

\setcounter{equation}{0}
\section{Generalized linear {\mathversion{bold}$r$}-matrix Poisson structure}
\label{Sect gen lin r-matrix}

Let now $\g$ have an additional structure of an associative algebra,
with the multiplication $(X,Y)\mapsto X\cdot Y$. The standard def\/inition of
the Lie bracket on $\g$ is then $[X,Y]=X\cdot Y-Y\cdot X$. Let $\g$ carry a 
{\it bi-invariant} scalar product, the property expressed 
by the equalities
\[
\langle X,Y\cdot Z\rangle=\langle X\cdot Y,Z\rangle=
\langle Z\cdot X,Y\rangle\qquad \forall \; X,Y,Z\in\g
\]
(this assures the invariance of the scalar product in the previous sense
relevant to the Lie algebra structure). We again identify $\g^*$ with $\g$ 
with the help of this scalar product.

Actually, there exist inf\/initely many ways to turn $\g$ into a Lie algebra, 
not only the standard one mentioned above. Indeed, f\/ix an arbitrary
element $\cF\in\g$, and def\/ine the corresponding Lie bracket as
\begin{equation}\label{other Lie br}
[X,Y]^{(\cF)}=X\cdot\cF\cdot Y-Y\cdot\cF\cdot X.
\end{equation}
We can replace the Lie bracket $[\cdot,\cdot]$ in all constructions of the
previous Section by the Lie bracket $[\cdot,\cdot]^{(\cF)}$, and come in this
way to the following def\/inition.
\begin{definition}
Let $\rR$ be a linear operator on $\g$. A 
{\bfseries \itshape  generalized linear {\mathversion{bold}$r$}-matrix
brac\-ket} on $\g$ corresponding to the operator $\rR$ and the element $\cF$ is
defined by:
\begin{equation}\label{lin br gen}
\{\varphi,\psi\}_1(L)=\frac{1}{2}\langle [\rR(\nabla\varphi(L)),
\nabla\psi(L)]^{(\cF)}+
[\nabla\varphi(L),\rR(\nabla\psi(L))]^{(\cF)},L\rangle.
\end{equation}
If this is indeed a Poisson bracket, it will be denoted by 
${\rm PB}_1(\rR;\cF)$. 
\end{definition}
Of course, a suf\/f\/icient condition for (\ref{lin br gen}) to be a Poisson 
bracket is that the following expression def\/ines a new structure of a Lie 
algebra on $\g$:
\begin{equation}\label{R algebra gen}
[X,Y]_{\rR}^{(\cF)}=\frac{1}{2}\left([\rR(X),Y]^{(\cF)}+
[X,\rR(Y)]^{(\cF)}\right).
\end{equation}
A suf\/f\/icient condition for this is given, clearly, by the
${\rm mYB}(\rR;\alpha)$, where the Lie bracket $[\cdot,\cdot]$ is replaced
through $[\cdot,\cdot]^{(\cF)}$, i.e.
\begin{equation}\label{mYB gen}
[\rR(X),\rR(Y)]^{(\cF)}-\rR\left([\rR(X),Y]^{(\cF)}+
[X,\rR(Y)]^{(\cF)}\right)=
-\alpha\,[X,Y]^{(\cF)}\  \ \forall \; X,Y\in \g.
\end{equation}
Notice that Ad-invariant functions $\psi$ of the Lie algebra 
$\left(\g,[\cdot,\cdot]^{(\cF)}\right)$ are characterized by the equation 
\begin{equation}\label{conj inv gen}
L\cdot\nabla\psi(L)\cdot\cF-\cF\cdot\nabla\psi(L)\cdot L=0\qquad 
\forall \; L\in\g.
\end{equation}
The analog of Theorem \ref{lin bracket Ham equations} can now be formulated.
\begin{theorem}\label{gen lin bracket Ham equations}
a) Let $\psi$ be an ${\rm Ad}$-invariant function of the Lie algebra 
$\left(\g,[\cdot,\cdot]^{(\cF)}\right)$. Then the Hamiltonian equation on 
$\g$ generated by the Hamilton function $\psi$ in the Poisson bracket 
${\rm PB}_1(\rR;\cF)$ reads:
\begin{equation}\label{lin br gen Lax}
\dot{L}=L\cdot C_2-C_1\cdot L,
\end{equation}
where
\begin{equation}\label{Cs in lin br gen Lax}
C_1=\frac{1}{2}\cF\cdot\rR(\nabla\psi(L)),\qquad 
C_2=\frac{1}{2}\rR(\nabla\psi(L))\cdot\cF.
\end{equation}

 b) ${\rm Ad}$-invariant functions of the algebra 
$\left(\g,[\cdot,\cdot]^{(\cF)}\right)$ are in involution with respect 
to the bracket ${\rm PB}_1(\rR;\cF)$.
\end{theorem}

The above notions simplify under some additional structural assumptions. 
Let $\g$ be an associative algebra with unit, and let $\cF\in\g$ be an 
{\it invertible element}. Then:
\begin{proposition}\label{lin bracket gen to usual}
a) The function $\psi$ is an ${\rm Ad}$-invariant of the algebra 
$\left(\g,[\cdot,\cdot]^{(\cF)}\right)$, i.e. {\rm(\ref{conj inv gen})} is
fulfilled, if and only if 
\[
\psi(L)=\varphi\left(L\cdot\cF^{-1}\right)=\varphi\left(\cF^{-1}\cdot L\right),
\]
where $\varphi$ is an ${\rm Ad}$-invariant of the algebra $\g$ with the 
standard Lie bracket. In this case
\[
\nabla\psi(L)=\cF^{-1}\cdot \nabla \varphi\left(L\cdot\cF^{-1}\right)=\nabla 
\varphi\left(\cF^{-1}\cdot L\right)\cdot\cF^{-1}.
\]

 b) The modified Yang-Baxter equation (\ref{mYB gen}) in the Lie 
algebra $\left(\g,[\cdot,\cdot]^{(\cF)}\right)$ for the operator $\rR$ is 
equivalent to the modified Yang--Baxter equation~(\ref{mYB}) in the 
algebra $\g$ with the standard Lie bracket for either of the two linear 
operators on $\g$,
\begin{equation}
\rR_1(X)=\cF\cdot\rR\left(\cF^{-1}\cdot X\right)\quad {and}\quad
\rR_2(X)=\rR\left(X\cdot\cF^{-1}\right)\cdot\cF.
\end{equation}

c) Under the map $L\mapsto L\cdot\cF^{-1}$ the bracket 
${\rm PB}_1(\rR;\cF)$ is pushed to ${\rm PB}_1(\rR_1)$, and under the map 
$L\mapsto\cF^{-1}\cdot L$ 
the bracket ${\rm PB}_1(\rR;\cF)$ is pushed to ${\rm PB}_1(\rR_2)$.
\end{proposition}

\setcounter{equation}{0}
\section{Quadratic {\mathversion{bold}$r$}-matrix Poisson structure}
\label{Sect quadratic r-mat br}

Let again $\g$ be an associative algebra with a nondegenerate invariant
scalar product. Let $\rA_1$, $\rA_2$, $\rS$ be three linear operators on 
$\g$, $\rA_1$ and $\rA_2$ being skew-symmetric:
\begin{equation}\label{skew}
\rA_1^*=-\rA_1,\qquad \rA_2^*=-\rA_2.
\end{equation}

\begin{definition} {\bf \cite{suris:S1}} A 
{\bfseries \itshape quadratic {\mathversion{bold}$r$}-matrix bracket} on
$\g$ corresponding to the triple $\rA_1$, $\rA_2$, $\rS$ is defined by:
\be
\ba{l}
\ds \{\varphi,\psi\}_2(L)  =  
\frac{1}{2}\langle\rA_1(d'\varphi(L)), d'\psi(L)\rangle
-\frac{1}{2}\langle\rA_2(d\varphi(L)),d\psi(L)\rangle 
\vspace{3mm}\\
\ds \qquad +\frac{1}{2}\langle\rS(d\varphi(L)),d'\psi(L)\rangle
-\frac{1}{2}\langle\rS^*(d'\varphi(L)),d\psi(L)\rangle, 
\ea \label{q br}
\ee
where we denote for brevity
\begin{equation}\label{dd'}
d\varphi(L)=L\cdot\nabla\varphi(L),\qquad 
d'\varphi(L)=\nabla\varphi(L)\cdot L.
\end{equation}
If this expression indeed defines a Poisson bracket, we shall denote 
it by ${\rm PB}_2(\rA_1,\rA_2,\rS)$.
\end{definition}
In what follows we usually suppose the following condition to be satisf\/ied:
\begin{equation}\label{sum}
\rA_1+\rS=\rA_2+\rS^*=\rR.
\end{equation}
Then a linearization of ${\rm PB}_2(\rA_1,\rA_2,\rS)$ in the unit element of
$\g$ coincides with ${\rm PB}_1(\rR)$, and we call the former a {\it
quadratization} of the latter.

\begin{theorem} {\bf \cite{suris:S1}} A sufficient condition for (\ref{q br})
to be a Poisson bracket is given by the equations (\ref{sum}) and
\begin{equation}\label{q br suf1}
{\rm mYB}(\rR;\alpha), \quad {\rm mYB}(\rA_1;\alpha),\quad 
{\rm mYB}(\rA_2;\alpha).
\end{equation}
Under these conditions the bracket ${\rm PB}_2(\rA_1,\rA_2,\rS)$ is 
compatible with ${\rm PB}_1(\rR)$.
\end{theorem}

If the operator $\rR$ is skew-symmetric and satisf\/ies ${\rm mYB}(\rR;\alpha)$,
then the Poisson bracket ${\rm PB}_2(\rR,\rR,0)$ is called {\it Sklyanin
bracket}~\cite{suris:STS}. The brackets ${\rm PB}_2(\rA,\rA,\rS)$ with a 
skew-symmetric operator $\rA$ and a symmetric operator $\rS$ were introduced
in~\cite{suris:LP,suris:OR}.

One of the most important properties of the $r$-matrix brackets is the 
following one.

\begin{theorem} {\bf \cite{suris:S1}}
Let the condition (\ref{sum}) be satisfied. Then:

a) Hamiltonian equations of motion on $\g$ corresponding to an 
${\rm Ad}$-invariant Hamilton function $\varphi$, have the Lax form
\begin{equation}\label{Lax with R q}
\dot{L}=[L,C],\qquad C=\frac{1}{2}\rR(d\varphi(L)).
\end{equation}

b) ${\rm Ad}$-invariant functions on $\g$ are in involution with
respect to ${\rm PB}_2(\rA_1,\rA_2,\rS)$. 
\end{theorem}

\setcounter{equation}{0}
\section{Quadratic {\mathversion{bold}$r$}-matrix 
structures on direct products}
\label{Sect quadr br on big algebra}

Quadratic $r$-matrix brackets have interesting and important features when
considered on a ``big'' algebra $\bg=\bigotimes\limits_{j=1}^n\g$. This algebra
carries a (nondegenerate, bi-invariant) scalar product 
\[
\langle\langle\mbL,\mbM\rangle\rangle=\sum_{k=1}^n\langle L_k,M_k\rangle.
\]
Working with linear operators on $\bg$, we use the following natural notations.
Let $\bA:\bg\mapsto\bg$ be a linear operator, let 
$\left(\bA(\bL)\right)_i$ be the $i$th component of $\bA(\bL)$; then we set
\begin{equation}\label{A i}
\left(\bA(\bL)\right)_i=\sum_{j=1}^n (\bA)_{ij}(L_j).
\end{equation}
For a smooth function $\Phi(\bL)$ on $\bg$ we also denote by $\nabla_j\Phi$, 
$d_j\Phi$, $d'_{j}\Phi$ the $j$th components of the corresponding objects.
We def\/ine also the {\it monodromy maps} $T_j:\bg\mapsto\g$, 
$1\le j\le n$, by the formula
\begin{equation}\label{Tj}
T_j(\mbL)=L_j\cdot\ldots\cdot L_1\cdot L_n\cdot\ldots\cdot L_{j+1}.
\end{equation}

Now let $\bA_1$, $\bA_2$, $\bS$ be 
linear operators on $\bg$ satisfying conditions
analogous to (\ref{skew}) and to (\ref{q br suf1}). One has, obviously:
\[
\left((\bA_1)_{ij}\right)^*=-(\bA_1)_{ji}, \qquad 
\left((\bA_2)_{ij}\right)^*=-(\bA_2)_{ji}, \qquad
\left((\bS)_{ij}\right)^*=(\bS^*)_{ji}.
\]
Then one can def\/ine the bracket ${\rm PB}_2(\bA_1,\bA_2,\bS)$ on $\bg$. In
components it reads:
\be\label{big br in components}
\ba{l}
\ds \{\Phi,\Psi\}_2(\mbL)  = 
\frac{1}{2}\sum_{i,j=1}^n\langle (\bA_1)_{ij}(d'_{j}\Phi),
d'_{i}\Psi\rangle-
\frac{1}{2}\sum_{i,j=1}^n\langle (\bA_2)_{ij}(d_j\Phi),
d_i\Psi\rangle
\vspace{3mm}\\
\ds \qquad +\frac{1}{2}\sum_{i,j=1}^n\langle (\bS)_{ij}(d_j\Phi),
d'_{i}\Psi\rangle
-\frac{1}{2}\sum_{i,j=1}^n\langle (\bS^*)_{ij}(d'_{j}\Phi),
d_i\Psi\rangle.
\ea
\ee

\begin{theorem}\label{monodromy} 
{\bf \cite{suris:S2}}
Let $\bg$ be equipped with the Poisson bracket ${\rm PB}_2(\bA_1,\bA_2,\bS)$.
Suppose that the following relations hold:
\begin{equation}\label{mono th sums}
(\bA_1)_{j+1,j+1}+(\bS)_{j+1,j}=(\bA_2)_{j,j}+(\bS^*)_{j,j+1}=\rR \quad
{\rm for\;\;all}\quad 1\le j\le n;
\end{equation}
\begin{equation}\label{mono th others}
(\bA_1)_{i+1,j+1}=-(\bS)_{i+1,j}=(\bS^*)_{i,j+1}=-(\bA_2)_{i,j}\quad
{\rm for}\quad i\neq j.
\end{equation}
Then each map $T_j:\bg\mapsto\g$ (\ref{Tj}) is Poisson, if the target 
space $\g$ is equipped with the Poisson bracket
\[
{\rm PB}_2
\left((\bA_1)_{j+1,j+1},(\bA_2)_{j,j},(\bS)_{j+1,j}\right).
\] 
Hamilton function of the form $\Phi(\mbL)=\varphi(L_n\cdot \ldots \cdot L_1)$, 
where $\varphi$ is an ${\rm Ad}$-invariant function on $\g$, generates 
Hamiltonian equations of motion on $\bg$ having the form of Lax triads:
\begin{equation}\label{big Lax with R}
\dot{L}_j=L_jC_{j-1}-C_j L_j,\qquad
C_j=\frac{1}{2}\rR(d\varphi(T_j)).
\end{equation}
(In all formulas the subscripts should be taken {\rm (mod $n$)}).
\end{theorem}

This theorem is a far-reaching
 generalization of the corresponding result for the
Sklyanin bracket ${\rm PB}_2(\bR,\bR,0)$ on $\bg$, which arises when $\bS=0$, 
$\bA_1=\bA_2=\bR$, and $(\bR)_{ij}=\rR\delta_{ij}$. In this case each map 
$T_j:\bg\mapsto\g$ {\rm(\ref{Tj})} is Poisson, if the target space $\g$ is 
equipped with the Sklyanin bracket ${\rm PB}_2(\rR,\rR,0)$~\cite{suris:STS}. 
Certain generalizations of the latter result appeared also, e.g.,
in \cite{suris:STS,suris:LP,suris:STSS}, 
but in all previously known formulations only few 
nonvanishing ``operator entries'' for the operators $\bA_1$, $\bA_2$, 
$\bS$ were allowed, namely ``diagonal'' ones for $\bA_1$, $\bA_2$, and 
``subdiagonal'' ones for $\bS$. In other words, all operators in 
(\ref{mono th others}) had to vanish. In the present paper, however, 
we shall need
this Theorem in its full generality (albeit only for $n=2$). For the f\/irst 
time this general bracket was applied in~\cite{suris:S1}.

We have discussed above the $r$-matrix origin of Lax equations, such as 
(\ref{Lax with R lin}), (\ref{Lax with R q}), or~(\ref{big Lax with R}).
If one is concerned with a Lax representation of a certain Hamiltonian f\/low
\begin{equation}\label{Ham syst}
\dot{x}=\{\rH,x\}
\end{equation}
on a Poisson manifold $\left(\cX,\{\cdot,\cdot\}\right)$, then f\/inding an 
$r$-matrix interpretation for it consists of f\/inding an $r$-matrix
bracket on $\g$ (or on $\bg$) such that the Lax matrix map $L:\cX\mapsto\g$ 
(resp. $\mbL:\cX\mapsto\bg$) is a Poisson map. Then the manifold consisting 
of the Lax matrices is a Poisson submanifold.

\setcounter{equation}{0}
\section{Notations}
\label{Sect notations}

\subsection{Algebras}
Two concrete algebras play the basic role in our presentation. 
They are well suited to describe various lattice systems with the so called 
open-end and periodic boundary conditions, respectively. Here are
the relevant def\/initions.

For the {\it open-end case} we always set $\g=gl(N)$, the algebra of
$N\times N$ matrices with the usual matrix product, the Lie bracket 
$[X,Y]=XY-YX$, and the nondegenerate bi-invariant scalar product
$\langle X,Y\rangle={\rm tr}(XY)$.  As a linear space, $\g$ may be 
represented as a direct sum
\[
\g=\bigoplus_{p=-N+1}^{N-1}\g_p,
\]
where the subspace $\g_p$ consists of matrices
\[
X=\sum_{j-k=p}x_{jk}E_{jk}.
\]
(Here and below $E_{jk}$ stands for the matrix whose only nonzero entry is on 
the intersection of the $j$th row and the $k$th column and is equal to~1).
The following sets are {\it subalgebras} of $\g$:
\[
\g_{>0}=\bigoplus_{p=1}^{N-1}\g_p,\qquad 
\g_{\ge 0}=\bigoplus_{p=0}^{N-1}\g_p,\qquad
\g_{<0}=\bigoplus_{p=-N+1}^{-1}\g_p,\qquad 
\g_{\le 0}=\bigoplus_{p=-N+1}^0\g_p,
\]
so that, for instance, $\g_{\ge 0}$ consists of lower triangular matrices,
$\g_{<0}$ consists of strictly upper triangular matrices, and $\g_0$ consists
of diagonal matrices. We shall always set
\[
\g_+=\g_{\ge 0},\qquad \g_-=\g_{<0}.
\]
Notice that, as a linear space,
\[
\g=\g_+\oplus\g_-,
\]

In the {\it periodic case} we always choose $\g$ as a certain {\it twisted 
loop algebra} over $gl(N)$. A loop algebra over $gl(N)$ is an algebra of 
Laurent polynomials with coef\/f\/icients from $gl(N)$ and a natural commutator
$\left[X\lambda^j,Y\lambda^k\right]=[X,Y]\lambda^{j+k}$. Our twisted algebra $\g$ is a 
subalgebra singled out by the additional condition
\[
\g=\left\{X(\lambda)\in gl(N)\left[\lambda,\lambda^{-1}\right]:
\Omega X(\lambda)\Omega^{-1}=X(\omega\lambda)\right\},
\]
where $\Omega={\rm diag}\left(1,\omega,\ldots,\omega^{N-1}\right)$, 
$\omega=\exp(2\pi i/N)$. In other words, elements of $\g$ satisfy
\begin{equation}
X(\lambda)=\sum_{p} \lambda^p
\sum_{j-k\equiv p \atop({\rm mod}\,N)}
 x_{jk}^{(p)}E_{jk}.
\end{equation}
The nondegenerate bi-invariant scalar product is chosen as 
\begin{equation}
\langle X(\lambda),Y(\lambda)\rangle={\rm tr}(X(\lambda)Y(\lambda))_0,
\end{equation}
the subscript 0 denoting the free term of the formal Laurent series. 

As a linear space, $\g$ is again a direct sum
\[
\g=\bigoplus _{p=-\infty}^{\infty}\g_p,
\]
where $\g_p$ consists of matrices
\[
X=\lambda^p\sum_{j-k\equiv p \atop({\rm mod}\,N)} x_{jk}E_{jk}.
\]
We have the subalgebras 
\[
\g_{>0}=\bigoplus_{p>0}\g_p,\qquad 
\g_{\ge 0}=\bigoplus_{p\ge 0}\g_p,\qquad
\g_{<0}=\bigoplus_{p<0}\g_p,\qquad 
\g_{\le 0}=\bigoplus_{p\le 0}\g_p.
\]
Again, we set 
\[
\g_+=\g_{\ge 0},\qquad \g_-=\g_{<0},
\]
so that, as a linear space,
\[
\g=\g_+\oplus\g_-.
\]

\subsection{Operators}

In both the open-end and the periodic cases we shall denote by
$\rP_{>0}$, $\rP_{\ge 0}$, etc. the projections from $\g$ onto the
corresponding subspace $\g_{>0}$, $\g_{\ge 0}$, etc., along the 
complementing subspace. For two such projections we use special notations:
\[
\pi_+=\rP_{\ge 0},\qquad \pi_-=\rP_{<0}.
\]
The basic operator governing the hierarchies of Lax equations, is
\begin{equation}\label{R}
\rR=\pi_+-\pi_-.
\end{equation}
An important property of this operator is given by the following general
statement:

\begin{theorem}\label{R split} {\bf \cite{suris:AM,suris:STS}} 
Let $\g$ as a linear space be a direct sum of its two subalgebras:
\[
\g=\g_+\oplus\g_-,
\]
and let $\pi_{\pm}$ denote projections from $\g$ onto $\g_{\pm}$ along the
complementary subspace. Then the operator
\[
\rR=\pi_+-\pi_-
\]
satisfies the modified Yang--Baxter equation ${\rm mYB}(\rR,\alpha)$ with
$\alpha=1$.
\end{theorem}
Let us remark that for the operators (\ref{R}) the Lax equations 
(\ref{Lax with R lin}), (\ref{Lax with R q}) are equivalent to
\begin{equation}\label{Tdot}
\dot{L}=[L,B]=[A,L],\quad{\rm where}\quad B=\pi_+(f(L)),\quad
 A=\pi_-(f(L))
\end{equation}
with $f(L)=\nabla\varphi(L)$ and $f(L)=d\varphi(L)$, respectively.
Similarly, the Lax equations on the ``big'' algebra $\bg$ 
(\ref{big Lax with R}) 
for the operator (\ref{R}) are equivalent to:
\begin{equation}\label{Tdot big}
\dot{L}_j=L_jB_{j-1}-B_jL_j=A_jL_j-L_jA_{j-1},
\ee
where
\[
B_j=\pi_+(f(T_j)),\qquad A_j=\pi_-(f(T_j)).
\]
We shall call such Lax equations {\it standard} ones. 

Denote by $\rR_0$, $\rP_0$ the skew-symmetric and the symmetric parts of the
operator (\ref{R}), respectively:
\[
\rR_0=(\rR-\rR^*)/2, \qquad \rP_0=(\rR+\rR^*)/2.
\]
It is easy to see that
\[
\rR_0=\rP_{>0}-\rP_{<0},
\]
and the notation $\rP_0$ agrees with the previously def\/ined projection to
$\g_0$: in the open-end case $\rP_0$ assigns to each matrix $X$ its 
diagonal part, and in the periodic case $\rP_0$ assigns to each Laurent 
series $X(\lambda)$ its free term. 

Let the skew-symmetric operator $\rW$ act on $\g_0$ according to
\begin{equation}\label{W}
\rW(E_{kk})=\sum_{j<k}E_{jj}-\sum_{j>k}E_{jj}=\sum_{j=1}^N w_{kj}E_{jj},
\end{equation}
where
\begin{equation}\label{w}
w_{kj}={\rm sgn}(k-j)=\left\{\begin{array}{rl}
1, & k>j\\ 0, & k=j  \\ -1, & k<j\end{array}\right.
\end{equation}
and extend $\rW$ on the rest of $\g$ according to $\rW=\rW\circ\rP_0$. 
Finally, def\/ine:
\begin{equation}\label{AS}
\rA_1=\rR_0+\rW,\quad \rA_2=\rR_0-\rW,\quad \rS=\rP_0-\rW,\quad
\rS^*=\rP_0+\rW.
\end{equation}
These operators will be basic building blocks in all quadratic 
$r$-matrix brackets appearing in this paper.


\setcounter{equation}{0}
\section{Multi-f\/ield analog of the Toda lattice}
\label{Sect GTL Lax}
The $(m+1)$-f\/ield phase space is, in the periodic case,
\begin{equation}\label{GTL phase sp}
\cT_{m+1}={\mathbb R}^{(m+1)N}\left(b,a^{(1)},\ldots,a^{(m)}\right),
\end{equation}
where the vectors
\[
b=(b_1,\ldots,b_N)\quad{\rm and}\quad 
a^{(j)}=\left(a_1^{(j)},\ldots,a_N^{(j)}\right)\quad(j=1,\ldots,m)
\]
represent the $m+1$ f\/ields. In the open-end
case the f\/ields $a^{(j)}$ consist of $N-j$ variables 
$\left(a_1^{(j)},\ldots,a_{N-j}^{(j)}\right)$ only.

Consider the Lax matrix
\begin{equation}\label{GTL T}
T\left(b,a^{(1)},\ldots,a^{(m)},\lambda\right)=
\lambda\sum_{k=1}^N E_{k+1,k}+\sum_{k=1}^N b_kE_{kk}+
\sum_{j=1}^m\lambda^{-j}\sum_{k=1}^N a_k^{(j)}E_{k,k+j}.
\end{equation}
The subspace of $\g$ consisting of all such matrices will be denoted
\begin{equation}\label{GTL T space}
\bT_{m+1}=\cE\oplus\bigoplus_{j=0}^m\g_{-j}.
\end{equation}
(We introduced here the notation $\cE=\lambda\sum\limits_{k=1}^N E_{k+1,k}$; 
recall that in the periodic case all subscripts are considered (mod~$N$), 
so that $E_{N+1,N}=E_{1,N})$. 

In the open-end case the Lax matrix, by convention, is given by
\[
T\left(b,a^{(1)},\ldots,a^{(m)}\right)=
\sum_{k=1}^{N-1} E_{k+1,k}+\sum_{k=1}^N b_kE_{kk}+
\sum_{j=1}^m\sum_{k=1}^{N-j} a_k^{(j)}E_{k,k+j},
\]
and the space of all such matrices is denoted still by (\ref{GTL T space}).
(So, in the open-end case $\cE=\sum\limits_{k=1}^{N-1}E_{k+1,k}$). In what 
follows we shall formulate the results and give the proofs mainly for the 
periodic case, since for the open-end case they appear to be parallel and 
only more simple.

\begin{proposition}\label{Lax for GTL} {\bf \cite{suris:K1}}
The Lax equation
\begin{equation}\label{GTL Lax}
\dot{T}=[T,B],
\end{equation}
with the Lax matrix (\ref{GTL T}) and the auxiliary matrix
\begin{equation}\label{GTL B}
B\left(b,a^{(1)},\ldots,a^{(m)},\lambda\right)=
\sum_{k=1}^N b_kE_{kk}+\lambda\sum_{k=1}^N E_{k+1,k},
\end{equation}
is equivalent to the following system of differential equations:
\begin{equation}\label{GTL}
\left\{\begin{array}{l}
\dot{b}_k=a_k^{(1)}-a_{k-1}^{(1)}, 
\vspace{2mm}\\ 
\ds \dot{a}_k^{(j)}=a_k^{(j)}\left(b_{k+j}-b_k\right)+
\left(a_k^{(j+1)}-a_{k-1}^{(j+1)}\right), 
\quad 1\le j\le m-1,
\vspace{2mm}\\ 
\dot{a}_k^{(m)}=a_k^{(m)}\left(b_{k+m}-b_k\right).
\end{array}\right.
\end{equation}
\end{proposition}
{\bf Proof} -- an elementary matrix calculation. \hfill \qed

\medskip

The system (\ref{GTL}) is what will be called ${\rm TL}_{m+1}$ -- the 
generalized $(m+1)$-f\/ield Toda lattice. It serves as a direct generalization 
of the usual Toda lattice TL. A consistent notation for TL would be 
${\rm TL}_2$, i.e. it corresponds to the value $m=1$.

\setcounter{equation}{0}
\section{Linear {\mathversion{bold}$r$}-matrix structure for
{\mathversion{bold}${\rm TL}_{m+1}$}}
\label{Sect GTL linear r-matrix}

Obviously, the matrix $B$ from (\ref{GTL B}) is nothing but the projection
\begin{equation}
B=\pi_+(T)
\end{equation}
in the standard decomposition of the algebra $\g$. So, the Lax representation 
of the system ${\rm TL}_{m+1}$ is of the standard form (\ref{Tdot}) with 
$f(T)=T$. We now give an $r$-matrix interpretation to this Lax representation,
def\/ining simultaneosly the whole hierarchy ${\rm TL}_{m+1}$. We start with the 
linear $r$-matrix Poisson structure on $\g$.

\begin{theorem}\label{Linear bracket for GTL} 
{\bf \cite{suris:STS,suris:RSTS}}
The set $\bT_{m+1}$ is a Poisson submanifold in the algebra $\g$ equipped
with the linear bracket ${\rm PB}_1(\rR)$.
\end{theorem}
{\bf Proof.} We have to demonstrate that at each point $T\in\bT_{m+1}$ 
every Hamiltonian vector f\/ield $\{\varphi,\cdot\}_1$ is tangent to $\bT_{m+1}$. 
To do this, notice that the value of the general Hamiltonian vector f\/ield
in the point $T$ is given by:
\[
2\{\varphi,T\}_1=[T,\rR(\nabla\varphi)]+\rR^*([T,\nabla\varphi]).
\]
Obviously, this can be represented in two equivalent forms:
\begin{equation}\label{GTL l proof aux2}
\{\varphi,T\}_1=[T,\rP_{\ge 0}(\nabla\varphi)]-\rP_{>0}([T,\nabla\varphi]),
\end{equation}
and
\begin{equation}\label{GTL l proof aux1}
\{\varphi,T\}_1=-[T,\rP_{<0}(\nabla\varphi)]+\rP_{\le 0}([T,\nabla\varphi]).
\end{equation}
Now take into account that $T\in\bigoplus\limits_{j=-m}^1\g_{j}$. 
Then, according to (\ref{GTL l proof aux2}), the value of $\{\varphi,T\}_1$ 
belongs to $\bigoplus\limits_{j\ge -m}\g_{j}$, and, according to 
(\ref{GTL l proof aux1}), this value belongs also to 
$\bigoplus\limits_{j\le 0}\g_{j}$. 
Hence it belongs to $\bigoplus\limits_{j=-m}^0\g_{j}$, 
which is the tangent space 
to the manifold $\bT_{m+1}$ in $\g$. \hfill \qed

\medskip

It remains to calculate the induced bracket on $\cT_{m+1}$. In what follows
we always def\/ine Poisson brackets by writing down all nonvanishing brackets
among the coordinate functions.
\begin{theorem} 
The coordinate representation of the restriction of ${\rm PB}_1(\rR)$ to
the subma\-ni\-fold $\bT_{m+1}$ is given by the formulas:
\begin{equation}\label{GTL l br b}
\left\{b_k,a_k^{(j)}\right\}_1=-a_k^{(j)},  \qquad 
\left\{a_k^{(j)},b_{k+j}\right\}_1=-a_k^{(j)},
\end{equation}
\begin{equation}\label{GTL l br a}
\left\{a_k^{(i)},a_{k+i}^{(j)}\right\}_1=-a_k^{(i+j)}  \quad (i+j\leq m).
\end{equation}
\end{theorem}
{\bf Proof.} We have, obviously:
\begin{equation}\label{GTL proof1 aux3}
\nabla b_k=\rR(\nabla b_k)=E_{kk},\qquad 
\nabla a_k^{(i)}=\rR\left(\nabla a_k^{(i)}\right)=\lambda^i E_{k+i,k}.
\end{equation}
Hence
\[
\ds \left\{b_k,a_{\ell}^{(j)}\right\}_1  =  
\langle T,\lambda^j[E_{kk},E_{\ell+j,\ell}]\rangle
  =  \langle T,\lambda^j E_{\ell+j,\ell}
(\delta_{k,\ell+j}-\delta_{k,\ell})\rangle
  =  a_{\ell}^{(j)}(\delta_{k,\ell+j}-\delta_{k,\ell}),
\]
which coincides with (\ref{GTL l br b}). Analogously,
\[
\ba{l}
\left\{a_k^{(i)},a_{\ell}^{(j)}\right\}_1  =  
\langle T,\lambda^{i+j}[E_{k+i,k},E_{\ell+j,\ell}]\rangle
 =  \langle T,\lambda^{i+j} E_{k+i,\ell}\delta_{k,\ell+j}
 -E_{\ell+j,k}\delta_{\ell,k+i}\rangle
\vspace{2mm}\\
\ds \qquad  = 
 a_{\ell}^{(i+j)}\delta_{k,\ell+j}-a_k^{(i+j)}\delta_{\ell,k+i},
\ea
\]
which coincides with (\ref{GTL l br a}). One shows similarly that the brackets
$\{b_k,b_j\}_1$ vanish. This f\/inishes the proof.\hfill \qed

\medskip

\noindent
{\bf Corollary.} {\bf \cite{suris:K1}}
{\it The system ${\rm TL}_{m+1}$ (\ref{GTL}) 
is Hamiltonian with respect to the bracket (\ref{GTL l br b}), 
(\ref{GTL l br a}), with the Hamilton function
\begin{equation}\label{GTL H2}
\rH_2=\frac{1}{2}\sum_{k=1}^N b_k^2+\sum_{k=1}^N a_k^{(1)}=
\frac{1}{2}\left({\rm tr}\left(T^2\right)\right)_0.
\end{equation}}

Theorem \ref{Linear bracket for GTL} not only provides us with
the Hamiltonian structure for the system ${\rm TL}_{m+1}$, but gives 
also an $r$-matrix explanation to the Lax representation from
Theorem \ref{Lax for GTL}, and allows to def\/ine the ${\rm TL}_{m+1}$
hierarchy as a set of f\/lows 
\[
\dot{T}=\left[T,\pi_+(\nabla\varphi(T))\right]
\]
with ${\rm Ad}$-invariant functions $\varphi$.

Also the complete integrability of ${\rm TL}_{m+1}$ with respect to the 
linear Poisson bracket can be studied on the base of Theorem \ref{Linear 
bracket for GTL}, because it guarantees involutivity of all spectral 
invariants of the matrix $T\left(b,a^{(1)},\ldots,a^{(m)},\lambda\right)$. 
Therefore, the complete integrability will follow as soon as it is proved
that the number 
of independent spectral invariants is large enough. This is almost obvious in
the case $m=1$, i.e. for the usual TL, but for $m>1$ it is no longer evident,
and, moreover, is no longer true in general. However, it may be shown that 
in the periodic case one gets the suf\/f\/icient number of independent spectral 
invariants. In the open-end case the situation is somewhat more delicate.
One has to introduce integrals which are not spectral invariants 
\cite{suris:DLNT,suris:EFS}.

\setcounter{equation}{0}
\section{Quadratic {\mathversion{bold}$r$}-matrix structure 
for {\mathversion{bold}${\rm TL}_{m+1}$}}
\label{Sect GTL quadratic r-matrix}

\begin{theorem}\label{Quadratic bracket for GTL} 
{\bf \cite{suris:S1,suris:O}}
The set $\bT_{m+1}$ is a Poisson submanifold in the algebra $\g$ equipped
with the quadratic bracket ${\rm PB}_2(\rA_1,\rA_2,\rS)$. 
\end{theorem}
{\bf Proof.} The Hamiltonian vector f\/ield on $\g$ 
with an arbitrary Hamilton function $\varphi$ with respect to the
bracket ${\rm PB}_2(\rA_1,\rA_2,\rS)$ is given by:
\begin{equation}\label{q br vf}
2\{\varphi,T\}_2 =T\cdot\rA_1(\nabla\varphi T)-\rA_2(T\nabla\varphi)\cdot T
+T\cdot\rS(T\nabla\varphi)-\rS^*(\nabla\varphi T)\cdot T.
\end{equation}
For the operators (\ref{AS}) this can be represented in two equivalent forms:
\be
\ba{l}
\ds 2\{\varphi,T\}_2  = 
2T\rP_{>0}(\nabla\varphi T)-2\rP_{>0}(T\nabla\varphi)T  
\vspace{2mm}\\
\ds \qquad  +T\rP_0(\nabla\varphi T)-\rP_0(T\nabla\varphi)T+
T\rP_0(T\nabla\varphi)-\rP_0(\nabla\varphi T)T   
\vspace{2mm}\\
\ds \qquad  +T\rW(\nabla\varphi T)+\rW(T\nabla\varphi)T
-T\rW(T\nabla\varphi)-\rW(\nabla\varphi T)T,
\ea \label{GTL proof2 vf1}
\ee
and
\be
\ba{l}
\ds 2\{\varphi,T\}_2  = 
-2T\rP_{<0}(\nabla\varphi T)+2\rP_{<0}(T\nabla\varphi)T   
\vspace{2mm}\\
\ds \qquad  -T\rP_0(\nabla\varphi T)+\rP_0(T\nabla\varphi)T+
T\rP_0(T\nabla\varphi)-\rP_0(\nabla\varphi T)T         
\vspace{2mm}\\
\ds \qquad  +T\rW(\nabla\varphi T)+\rW(T\nabla\varphi)T
-T\rW(T\nabla\varphi)-\rW(\nabla\varphi T)T.
\ea\label{GTL proof2 vf2}
\ee
The f\/irst expression assures that for $T\in\bT_{m+1}$ this vector f\/ield 
takes the value in $\bigoplus\limits_{j\ge -m}\g_j$, 
and the second expression yields 
that that this vector f\/ield belongs also to $\bigoplus\limits_{j\le 1}\g_j$. 
Hence it belongs to ${\mathop {\oplus}\limits_{j=-m}^1}\g_j$. 
It remains to prove that the 
$\g_1$ component of this vector f\/ield vanishes. Up to now everything held
true for an arbitrary operator $\rW$ with the values in $\g_0$. Now the
concrete expression (\ref{W}) for $\rW$ becomes crucial. From~(\ref{GTL T}),
(\ref{GTL proof2 vf2}) we have the following expression for the component in
question:
\be
\ba{l}
2\rP_1\left(\{\varphi,T\}_2\right) =  
-{\mathcal E}\rP_0(\nabla\varphi T)+\rP_0(T\nabla\varphi){\mathcal E}+
{\mathcal E}\rP_0(T\nabla\varphi)-\rP_0(\nabla\varphi T) {\mathcal E} 
\vspace{2mm}\\
\ds \qquad +{\mathcal E}\rW(\nabla\varphi T)+\rW(T\nabla\varphi){\mathcal E}
-{\mathcal E}\rW(T\nabla\varphi)-\rW(\nabla\varphi T){\mathcal E} .
\ea\label{TL proof2 aux3}
\ee

Due to $\rW=\rW\circ \rP_0$ the condition for (\ref{TL proof2 aux3}) to 
vanish may be presented as
\begin{equation}\label{TL proof2 aux4}
\rW(D){\mathcal E}-{\mathcal E}\rW(D)+D{\mathcal E}+{\mathcal E}D=
\rW(D'){\mathcal E}-{\mathcal E}\rW(D')+D'{\mathcal E}+{\mathcal E}D',
\end{equation}
where
\begin{equation}\label{TL proof2 aux5}
D=\rP_0(T\nabla\varphi),\qquad D'=\rP_0(\nabla\varphi T).
\end{equation}
Now a direct calculation shows that for the operator $\rW$ in (\ref{W})
and for an arbitrary diagonal matrix $D$ one has:
\begin{equation}\label{TL proof2 aux6}
\rW(D){\mathcal E}-{\mathcal E}\rW(D)+D{\mathcal E}+{\mathcal E}D=
\left\{\begin{array}{cl}
2\lambda\,{\rm tr}(D)\,E_{1,N},  & {\rm periodic\ case}
\vspace{2mm}\\
0, & {\rm open-end\ case}
\end{array}\right.
\end{equation}
This proves (\ref{TL proof2 aux4}) for the matrices (\ref{TL proof2 aux5}),
since ${\rm tr}(D)={\rm tr}(T\nabla\varphi)={\rm tr}(\nabla\varphi T)=
{\rm tr}(D')$ (note that in the open-end case the last argumentation is
superf\/luos). \hfill\qed

\medskip

Theorem \ref{Quadratic bracket for GTL} gives the second Hamiltonian 
interpretation of the system ${\rm TL}_{m+1}$, and yields simultaneously 
the complete integrability of ${\rm TL}_{m+1}$ in this Hamiltonian 
formulation (at least in the periodic case). Moreover, it delivers an 
alternative $r$-matrix explaination of the Lax representation from 
Theorem~\ref{Lax for GTL}.
It remains to calculate the coordinate representation of the restriction
of ${\rm PB}_2(\rA_1,\rA_2,\rS)$ to $\bT_{m+1}$. In the sequel we often 
use in our formulas the following notational convention: 
\begin{equation}\label{GTL convention}
a_k^{(0)}=b_k; \qquad a_k^{(i)}=0\quad  {\rm for}\quad  i<0\quad 
{\rm or}\quad  i>m,
\end{equation}
whenever applicable.

\begin{theorem}\label{Quadratic bracket for GTL coord} 
The bracket induced on $\bT_{m+1}$ by ${\rm PB}_2(\rA_1,\rA_2,\rS)$ is given 
by the following formulas (in (\ref{GTL q br aa1}) we suppose that $i<j$, 
and in (\ref{GTL q br aa2}), (\ref{GTL q br aa3}) we suppose that 
$i\le j$):
\be
\left\{b_k,b_{k+1}\right\}_2  =  -a_k^{(1)},  
\label{GTL q br bb0}
\ee
\be
\left\{b_k,a_{k+1}^{(j)}\right\}_2  =  -a_k^{(j+1)},  \qquad  
\left\{a_k^{(j)},b_{k+j+1}\right\}_2=-a_k^{(j+1)}, 
\label{GTL q br ba0}
\ee
\be
\left\{b_k,a_k^{(j)}\right\}_2  =  -b_ka_k^{(j)},  \qquad  
\left\{a_k^{(j)},b_{k+j}\right\}_2=-a_k^{(j)}b_{k+j}, 
\label{GTL q br ba1}
\ee
\be
\left\{a_k^{(i)},a_{k+i+1}^{(j)}\right\}_2  =  -a_k^{(i+j+1)},
\label{GTL q br aa0}
\ee
\be
\left\{a_k^{(i)},a_k^{(j)}\right\}_2  =  -a_k^{(i)}a_k^{(j)},  \qquad  
\left\{a_k^{(j)},a_{k+j-i}^{(i)}\right\}_2=-a_k^{(j)}a_{k+j-i}^{(i)},
\label{GTL q br aa1}
\ee
\be
\left\{a_k^{(i)},a_{k+\ell}^{(j)}\right\}_2  =  -a_k^{(i)}a_{k+\ell}^{(j)}
-a_k^{(j+\ell)}a_{k+\ell}^{(i-\ell)} \qquad (1\le\ell\le i),   
\label{GTL q br aa2}
\ee
\be
\left\{a_k^{(j)},a_{k+\ell}^{(i)}\right\}_2  =  -a_k^{(j)}a_{k+\ell}^{(i)}
-a_k^{(i+\ell)}a_{k+\ell}^{(j-\ell)} \qquad (j-i+1\le\ell\le j).   
\label{GTL q br aa3}
\ee
\end{theorem}
{\bf Proof.} Notice that without the convention (\ref{GTL convention}) the
second term on the right-hand side of (\ref{GTL q br aa2}) would have to
be multiplied by $\chi_{\ell}\left(1,\min(i,m-j)\right)$, and, likewise,
the second term on the right-hand side of (\ref{GTL q br aa3}) would have to
be multiplied by $\chi_{\ell}\left(j-i+1,\min(j,m-i)\right)$, where
\[
\chi_{\ell}(\alpha,\beta)=\left\{\begin{array}{l}
1,\quad \alpha\le\ell\le\beta
\vspace{1mm}\\ 0,\quad {\rm else}
\end{array}\right.
\]
is the characteristic function of the interval $[\alpha,\beta]$.
Notice also that under the convention~(\ref{GTL convention}) the  
formulas (\ref{GTL q br bb0}), (\ref{GTL q br ba0}) might be seen 
as particular cases of (\ref{GTL q br aa0}). Similarly, (\ref{GTL q br ba1}) 
might be seen as a particular case of (\ref{GTL q br aa1}).

Toward the calculation of the induced bracket, we use the def\/ining formula
\begin{eqnarray}\label{GTL proof2 PB}
\lefteqn{
2\{\varphi,\psi\}_2 =
2\langle\rP_{>0}(\nabla\varphi T),\nabla\psi T\rangle
-2\langle\rP_{>0}(T\nabla\varphi),T\nabla\psi\rangle}\nonumber\\ 
&& \hspace*{-5pt}+\langle\rP_0(\nabla\varphi T),\nabla\psi T\rangle
-\langle\rP_0(T\nabla\varphi),T\nabla\psi\rangle 
+\langle\rP_0(T\nabla\varphi),\nabla\psi T\rangle-
\langle\rP_0(\nabla\varphi T),T\nabla\psi\rangle \nonumber\\
&& \hspace*{-5pt}+\langle\rW(\nabla\varphi T),\nabla\psi T\rangle+
\langle\rW(T\nabla\varphi),T\nabla\psi\rangle       
-\langle \rW(T\nabla\varphi),\nabla\psi T\rangle-
\langle\rW(\nabla\varphi T),T\nabla\psi\rangle. \nonumber\\
\end{eqnarray}
Take here
\[
\varphi(T)=a_k^{(i)},\qquad \psi(T)=a_{k+\ell}^{(j)},
\]
so that
\[
\nabla\varphi=\lambda^iE_{k+i,i},\qquad
\nabla\psi=\lambda^jE_{k+\ell+j,k+\ell}.
\]
Further, we f\/ind:
\begin{equation}
\label{GTL proof2 aux1}
\nabla\varphi T=\lambda^{i+1}E_{k+i,k-1}+
\sum_{\beta=0}^m\lambda^{i-\beta}a_k^{(\beta)}E_{k+i,k+\beta},
\end{equation}
\begin{equation}\label{GTL proof2 aux2}
T\nabla\varphi=\lambda^{i+1}E_{k+i+1,k}+
\sum_{\beta=0}^m\lambda^{i-\beta}a_{k+i-\beta}^{(\beta)}E_{k+i-\beta,k}.
\end{equation}
We consider f\/irst the contribution to the Poisson bracket 
$\{\varphi,\psi\}_2=
\left\{a_k^{(i)},a_{k+\ell}^{(j)}\right\}_2$ from the f\/irst line in (\ref{GTL
proof2 PB}):
\[
\ba{l}
\ds \left\langle \lambda^{i+1}E_{k+i,k-1}+\sum_{\beta=0}^{i-1}\lambda^{i-\beta}
a_k^{(\beta)}E_{k+i,k+\beta},
\sum_{\gamma=j+1}^m\lambda^{j-\gamma}a_{k+\ell}^{(\gamma)}
E_{k+\ell+j,k+\ell+\gamma}\right\rangle 
\vspace{3mm}\\
\ds \quad
-\left\langle \lambda^{i+1}E_{k+i+1,k}+\sum_{\beta=0}^{i-1}\lambda^{i-\beta}
a_{k+i-\beta}^{(\beta)}E_{k+i-\beta,k},
\sum_{\gamma=j+1}^m\lambda^{j-\gamma}a_{k+\ell+j-\gamma}^{(\gamma)}
E_{k+\ell+j-\gamma,k+\ell}\right\rangle.
\ea 
\]
Calculating these scalar products, we f\/ind:
\[
\ba{l}
\ds  = \sum_{\gamma=j+1}^m\left(
a_{k+\ell}^{(\gamma)}\delta_{-1,\ell+j}\delta_{i,\ell+\gamma}-
a_{k+\ell+j-\gamma}^{(\gamma)}
\delta_{0,\ell+j-\gamma}\delta_{i+1,\ell}\right)
\vspace{3mm}\\
\ds \qquad +\sum_{\beta=0}^{i-1}\sum_{\gamma=j+1}^m\left(
a_k^{(\beta)}a_{k+\ell}^{(\gamma)}\delta_{\beta,\ell+j}\delta_{i,\ell+\gamma}
-a_{k+i-\beta}^{(\beta)}a_{k+\ell+j-\gamma}^{(\gamma)}\delta_{0,\ell+j-\gamma}
\delta_{i-\beta,\ell}\right)
\vspace{3mm}\\
\ds \qquad
 =  a_{k+\ell}^{(i+j+1)}\,\delta_{\ell,-j-1}-a_k^{(i+j+1)}\delta_{\ell,i+1}
+\varkappa_{k\ell}^{(ij)}a_k^{(\ell+j)}a_{k+\ell}^{(i-\ell)},
\ea
\]
where
\be
\ba{l}
\ds \varkappa_{k\ell}^{(ij)}  =  
\chi_{\ell}(-j,i-j-1)\chi_{\ell}(i-m,i-j-1)
-\chi_{\ell}(1,m-j)\chi_{\ell}(1,i) 
\vspace{2mm}\\
\ds \qquad  =  \chi_{\ell}\left(-\min(j,m-i),i-j-1\right)
-\chi_{\ell}\left(1,\min(i,m-j)\right).
\ea
\ee
Assuming, for the sake of def\/initeness, that $i\le j$, we see that the 
intervals of the two characteristic functions in the last line do not
intersect. So, we found the contributions to the Poisson bracket
$\left\{a_k^{(i)},a_{k+\ell}^{(j)}\right\}_2$ described by the formulas
(\ref{GTL q br bb0}), (\ref{GTL q br ba0}), (\ref{GTL q br aa0}), and the
second terms on the right-hand sides of (\ref{GTL q br aa2}),
(\ref{GTL q br aa3}). Now we turn to the contribution of the remaining
part of (\ref{GTL proof2 PB}). To this end, we notice that
\[
\rP_0(\nabla\varphi T)=a_k^{(i)}E_{k+i,k+i},\qquad
\rP_0(T\nabla\varphi)=a_k^{(i)}E_{kk}.
\]
Using also (\ref{W}), we f\/ind that the contribution under consideration 
is equal to
\begin{equation}\label{GTL q br aux2}
\varepsilon_{k\ell}^{(ij)}a_k^{(i)}a_{k+\ell}^{(j)},
\end{equation}
where
\begin{equation}\label{GTL q br aux3}
\ba{l}
\ds \varepsilon_{k\ell}^{(ij)}=\frac{1}{2}
(\delta_{i,\ell+j}-\delta_{0,\ell}
+\delta_{0,\ell+j}-\delta_{i,\ell}
\vspace{2mm}\\
\ds \qquad +w_{k+i,k+\ell+j}+w_{k,k+\ell}-
w_{k,k+\ell+j}-w_{k+i,k+\ell}).
\ea
\end{equation}
A direct analysis allows to f\/ind a compact expression for this coef\/f\/icient.
In the case $i\le j$ we have:
\begin{equation}\label{GTL q br aux4}
\varepsilon_{k\ell}^{(ij)}=\chi_{\ell}(-j,i-j)-\chi_{\ell}(0,i).
\end{equation}
Again, by $i<j$ the intervals of these two characteristic functions 
do not intersect, and by $i=j$ their only common point is $\ell=0$,
so that the last formula exactly describes the contribution 
into the Poisson bracket made by (\ref{GTL q br ba1}), (\ref{GTL q br aa1}), 
and the f\/irst terms on the right-hand sides of (\ref{GTL q br aa2}), 
(\ref{GTL q br aa3}). This f\/inishes the proof. \hfill \qed

\medskip

\noindent
{\bf Corollary.}  {\bf \cite{suris:K1}}
{\it The system ${\rm TL}_{m+1}$ (\ref{GTL})
is Hamiltonian with respect to the bracket given in 
Theorem~{\rm\ref{Quadratic bracket for GTL coord}}, with the Hamilton function
\begin{equation}\label{GTL H1}
\rH_1=\sum_{k=1}^N b_k=\left({\rm tr}(T)\right)_0.
\end{equation}}

\setcounter{equation}{0}
\section[Example: ${\rm TL}_3$]
{Example: {\mathversion{bold}${\rm TL}_3$},
 \\ the three-f\/ield analog of the Toda lattice}
\label{Sect GTL example}

In order to illustrate the above results, we give their specialization
for the case next in complexity after the usual Toda lattice TL, i.e. for
$m=2$. The equations of motion of the system ${\rm TL}_3$ read:
\begin{equation}\label{GTL3}
\left\{\begin{array}{l}
\dot{b}_k=a_k^{(1)}-a_{k-1}^{(1)}, 
\vspace{2mm}\\ 
\dot{a}_k^{(1)}=a_k^{(1)}\left(b_{k+1}-b_k\right)+
\left(a_k^{(2)}-a_{k-1}^{(2)}\right), 
\vspace{2mm}\\ 
\dot{a}_k^{(2)}=a_k^{(2)}\left(b_{k+2}-b_k\right).
\end{array}\right.
\end{equation}
The linear invariant Poisson bracket of this system is given by:
\be
\ba{l}
\left\{b_k,a_k^{(1)}\right\}_1=-a_k^{(1)},  \qquad 
\left\{a_k^{(1)},b_{k+1}\right\}_1=-a_k^{(1)}, 
\vspace{2mm}\\
\left\{b_k,a_k^{(2)}\right\}_1=-a_k^{(2)},  \qquad 
\left\{a_k^{(2)},b_{k+2}\right\}_1=-a_k^{(2)}, \label{GTL3 l br}
\qquad  
\left\{a_k^{(1)},a_{k+1}^{(1)}\right\}_1=-a_k^{(2)}. 
\ea
\ee
The quadratic invariant Poisson bracket, compatible with the previous one,
is given by:
\be
\ba{l}
\left\{b_k,b_{k+1}\right\}_2  =  -a_k^{(1)},  
\qquad
\left\{b_k,a_k^{(1)}\right\}_2  =  -b_ka_k^{(1)},  \qquad  
\left\{a_k^{(1)},b_{k+1}\right\}_2=-a_k^{(1)}b_{k+1}, 
\vspace{2mm}\\
\left\{b_k,a_{k+1}^{(1)}\right\}_2  =  -a_k^{(2)},  \qquad  
\left\{a_k^{(1)},b_{k+2}\right\}_2=-a_k^{(2)}, 
\qquad \left\{b_k,a_k^{(2)}\right\}_2  =  -b_ka_k^{(2)},  
\vspace{2mm}\\
\left\{a_k^{(2)},b_{k+2}\right\}_2=-a_k^{(2)}b_{k+2}, 
\qquad
\left\{a_k^{(1)},a_{k+1}^{(1)}\right\}_2  =  -a_k^{(1)}a_{k+1}^{(1)}-
a_k^{(2)}b_{k+2}, \label{GTL3 q br}
\vspace{2mm}\\ 
\left\{a_k^{(1)},a_k^{(2)}\right\}_2  =  -a_k^{(1)}a_k^{(2)},  \qquad  
\left\{a_k^{(2)},a_{k+1}^{(1)}\right\}_2=-a_k^{(2)}a_{k+1}^{(1)},
\vspace{2mm}\\
\left\{a_k^{(1)},a_{k+1}^{(2)}\right\}_2  = 
 -a_k^{(1)}a_{k+1}^{(2)},  \qquad  
\left\{a_k^{(2)},a_{k+2}^{(1)}\right\}_2=-a_k^{(2)}a_{k+2}^{(1)},
\vspace{2mm}\\
\left\{a_k^{(2)},a_{k+1}^{(2)}\right\}_2  =  -a_k^{(2)}a_{k+1}^{(2)},  
\qquad \left\{a_k^{(2)},a_{k+2}^{(2)}\right\}_2  =  -a_k^{(2)}a_{k+2}^{(2)}. 
\ea\hspace{-5.86pt}
\ee

\setcounter{equation}{0}
\section[Multi-f\/ield RTL: f\/irst version]{Multi-f\/ield analog of the 
relativistic \\ Toda lattice: the f\/irst construction}
\label{Sect alternative GRTL}

The essence of the relativistic ansatz of \cite{suris:GK} 
consists in multiplying the nonrelativistic Lax matrix by
$\cF^{-1}(\lambda)$, where
\begin{equation}\label{GRTL GK F}
\cF(\lambda)=I-\alpha\lambda\sum_{k=1}^N E_{k+1,k}\,,
\end{equation}
and the (small) parameter $\alpha$ is the relativistic parameter 
(the inverse speed of light). So the Lax matrix of the relativistic 
analog of ${\rm TL}_{m+1}$ is
\[
T\cF^{-1}\quad {\rm or}\quad \cF^{-1}T,
\]
where $T$ is the matrix from (\ref{GTL T}). It will be convenint to
consider also the matrices
\be
\ba{l}
\cT_1\left(b,a^{(1)},\ldots,a^{(m)},\lambda\right)  = 
I+\alpha T\left(b,a^{(1)},\ldots,a^{(m)},\lambda\right)\cdot\cF^{-1}(\lambda)
\vspace{2mm}\\
 \qquad 
=  \cL\left(b,a^{(1)},\ldots,a^{(m)},\lambda)\cdot\cF^{-1}(\lambda\right),
\ea\label{GRTL GK T1}
\ee
\be
\ba{l}
\ds \cT_2\left(b,a^{(1)},\ldots,a^{(m)},\lambda\right)  =  
I+\alpha\cF^{-1}(\lambda)\cdot T\left(b,a^{(1)},\ldots,a^{(m)},\lambda\right)
\vspace{2mm}\\
 \qquad  
=  \cF^{-1}(\lambda)\cdot
\cL\left(b,a^{(1)},\ldots,a^{(m)},\lambda\right),
\ea\label{GRTL GK T2}
\ee
where 
\be\label{GRTL GK L}
\ba{l}
\cL\left(b,a^{(1)},\ldots,a^{(m)},\lambda\right)  =  \cF(\lambda)+\alpha 
T\left(b,a^{(1)},\ldots,a^{(m)},\lambda\right) 
\vspace{3mm}\\
\ds\qquad  =  \sum_{k=1}^N (1+\alpha b_k)E_{kk}
+\alpha\sum_{j=1}^m\lambda^{-j} \sum_{k=1}^N a_k^{(j)}E_{k,k+j}.
\ea
\ee

The arising hierarchy, as well as its simplest 
(``minus f\/irst'') f\/low will be denoted ${\rm RTL}_{m+1}^{(-)}(\alpha)$.
We shall use the notation
\begin{equation}\label{GRTL alt phase sp}
\cR\cT_{m+1}^{(-)}={\mathbb R}^{(m+1)N}\left(b,a^{(1)},\ldots,a^{(m)}\right)
\end{equation}
for the phase variables of this hierarchy.

\begin{proposition}\label{Lax for GRTL of GK} 
{\bf \cite{suris:K3}}
Consider the matrix differential equation
\begin{equation}\label{GRTL GK Lax L}
\dot{T}=T\cB_2-\cB_1T,
\end{equation}
with the auxiliary matrices
\be
\cB_1  =  \sum_{k=1}^N\frac{b_k}{1+\alpha b_k}E_{kk}+\lambda
\sum_{k=1}^N \frac{1}{1+\alpha b_k}\,E_{k+1,k}, \label{GRTL GK B1}
\ee
\be
\cB_2  =  \sum_{k=1}^N\frac{b_k}{1+\alpha b_k}\,E_{kk}+\lambda
\sum_{k=1}^N \frac{1}{1+\alpha b_{k+1}}\,E_{k+1,k}. \label{GRTL GK B2}
\ee
The matrices $\cB_i$ satisfy the relation
\begin{equation}\label{GRTL GK Lax F}
\cF\cB_2-\cB_1\cF=0,
\end{equation}
which assures that the equation {\rm(\ref{GRTL GK Lax L})} is equivalent to 
either of the following two Lax equations:
\begin{equation}\label{GRTL GK Lax}
\dot{\cT}_i=[\cT_i,\cB_i],\qquad i=1,2.
\end{equation}
These matrix differential equations are equivalent to the following system:
\begin{equation}\label{GRTL GK}
\left\{\begin{array}{l}
\ds \dot{b}_k=\frac{a_k^{(1)}}{1+\alpha b_{k+1}}-
\frac{a_{k-1}^{(1)}}{1+\alpha b_{k-1}},
\vspace{3mm}\\
\ds \dot{a}_k^{(j)}=a_k^{(j)}\left(\frac{b_{k+j}}{1+\alpha b_{k+j}}
-\frac{b_k}{1+\alpha b_k}\right)+
\left(\frac{a_k^{(j+1)}}{1+\alpha b_{k+j+1}}
-\frac{a_{k-1}^{(j+1)}}{1+\alpha b_{k-1}}\right),
\vspace{3mm}\\
\ds \dot{a}_k^{(m)}=a_k^{(m)}\left(\frac{b_{k+m}}{1+\alpha b_{k+m}}
-\frac{b_k}{1+\alpha b_k}\right), \qquad 1\le j\le m-1.
\end{array}\right.
\end{equation}
\end{proposition}

In \cite{suris:K3}
this system was called ``the shadow relativistic lattice KP''.
Obviously, the system (\ref{GRTL GK}) is a one-parameter perturbation 
of (\ref{GTL}). In order to def\/ine the hierarchy attached to the f\/low 
${\rm RTL}_{m+1}^{(-)}(\alpha)$ and to f\/ind its Hamiltonian structure, 
we shall use the $r$-matrix theory.

\setcounter{equation}{0}
\section{Linear {\mathversion{bold}$r$}-matrix 
structure for {\mathversion{bold}${\rm RTL}_{m+1}^{(-)}(\alpha)$}}
\label{Sect r-matrix for GRTL of GK}

Let us consider the generalized $r$-matrix bracket ${\rm PB}_1(\rR;\cF)$ 
introduced in Sect.~\ref{Sect gen lin r-matrix}, with the standard operator
\begin{equation}\label{RTL R}
\rR=\pi_+-\pi_-.
\end{equation}
First of all, we have to be sure that this is indeed a Poisson bracket. 
Notice that the matrix $\cF$ is invertible. Therefore the constructions of 
Proposition~\ref{lin bracket gen to usual} are applicable.
\begin{lemma}\label{MYB for RTL alternative Lax}
Let the linear operators $\rR_{1,2}$ on $\g$ be defined by the formulas
\begin{equation}\label{RTL alt Lax R}
\rR_1(X)=\cF\cdot\rR\left(\cF^{-1}X\right),\qquad 
\rR_2(X)=\rR\left(X\cF^{-1}\right)\cdot\cF.
\end{equation}
Then these operators satisfy the modified Yang--Baxter equation.
\end{lemma}
{\bf Proof.} The proofs for $\rR_1$ and $\rR_2$ are similar, therefore
we restrict ourselves to the $\rR_1$ case. We prove that this operator
can be represented as
\begin{equation}\label{R for RTL alternative Lax}
\rR_1=\rP_{(+)}-\rP_{(-)},
\end{equation}
where $\rP_{(\pm)}$ denote the projections from $\g$ to the corresponding 
subspaces $\g_{(\pm)}$ in a certain decomposition of $\g$ (as a linear space) 
into a direct sum
\begin{equation}\label{split for RTL alternative Lax}
\g=\g_{(+)}\oplus\g_{(-)},
\end{equation}
each one of the subspaces $\g_{(\pm)}$ being also a subalgebra and a Lie
subalgebra of $\g$. Then the statement of the Lemma is assured by the 
general construction of Theorem \ref{R split}. To prove the above claim, set
\[
\g_{(+)}=\cF\g_+,\qquad \g_{(-)}=\cF\g_-,
\] 
and 
\[
\rP_{(+)}(X)=\cF\pi_+\left(\cF^{-1}X\right),
\qquad \rP_{(-)}(X)=\cF\pi_-\left(\cF^{-1}X\right).
\]
Obviously, from these def\/initions there follows the direct decomposition
(\ref{split for RTL alternative Lax}), the representation
(\ref{R for RTL alternative Lax}), as well as the following statements:
\[
\rP_{(+)}(X)+\rP_{(-)}(X)=X,\qquad
\rP_{(+)}(X)\in\g_{(+)},\qquad \rP_{(-)}(X)\in\g_{(-)}.
\]
It remains to show that $\g_{(\pm)}$ are subalgebras. But this follows
from the obvious relations
\[
X\cF Y\in\g_+\quad \forall \; X,Y\in\g_+ \quad \Rightarrow \quad
\cF X\cdot\cF Y\in\g_{(+)}\quad\forall\; \cF X,\cF Y\in\g_{(+)},
\]
and
\[
X\cF Y\in\g_-\quad \forall \; X,Y\in\g_- \quad \Rightarrow \quad
\cF X\cdot\cF Y\in\g_{(-)}\quad \forall\; \cF X,\cF Y\in\g_{(-)}.
\]
The Lemma is proved.\hfill \qed

\medskip

\noindent
{\bf Remark.} It is important to notice that the operators $\rR_1$ and $\rR_2$
correspond to {\it different} splittings (\ref{split for RTL alternative Lax}).
In both cases one has $\g_{(+)}=\g_+=\g_{\ge 0}$, but $\g_{(-)}$ are 
dif\/ferent subalgebras of $\g_{\le 0}$ for the two cases. Let us give also 
somewhat more explicit formulas for the operators~$\rR_{1,2}$. 
We use the following notation for the negative part of $X$:
\begin{equation}\label{to GK symbol}
\pi_-(X)=\sum_{\ell>0}\lambda^{-\ell}\sum_{k=1}^N x_k^{(-\ell)}E_{k,k+\ell}.
\end{equation}
Then for the operators $\rR_{1,2}=\rP_{(+)}-\rP_{(-)}$ we have:
\begin{equation}\label{R12 splitting}
\rP_{(+)}(X)=\pi_+(X)+\sigma_{1,2}(X),\qquad
\rP_{(-)}(X)=\pi_-(X)-\sigma_{1,2}(X),
\end{equation}
respectively, where the operators $\sigma_{1,2}:\g\mapsto\g_0$ act according 
to the formula
\begin{equation}\label{sigma12}
\sigma_1(X)=\sum_{\ell>0}\alpha^{\ell}\sum_{k=1}^N x_k^{(-\ell)}
E_{k+\ell,k+\ell},\qquad
\sigma_2(X)=\sum_{\ell>0}\alpha^{\ell}\sum_{k=1}^N x_k^{(-\ell)}E_{kk}.
\end{equation}
In words: the operators $\sigma_{1,2}$ assign the weights $\alpha^{\ell}$ to
the elements of $\g_{-\ell}$, and shift all entries of $\pi_-(X)$ to the
diagonal positions: $\sigma_1$ along the rows where they stand, and 
$\sigma_2$ along the columns where they stand. In this form the operators
$\sigma_{1,2}$ appeared in \cite{suris:OR,suris:GK}
(in the latter reference -- under the
name ``relativistic symbol'').

\begin{theorem}\label{linear r-matrix for GRTL of GK}
a) The set $\bT_{m+1}$ is a Poisson submanifold in $\g$ equipped
with the linear $r$-matrix bracket ${\rm PB}_1(\rR;\cF)$. 

b) Let $\varphi$ be an {\rm Ad}-invariant function on $\g$, 
and let $\psi(T)=\varphi\left(T\cF^{-1}\right)=\varphi\left(\cF^{-1}T\right)$.
Then the Hamiltonian equation on $\g$ with respect to ${\rm PB}_1(\rR,\cF)$ 
with the Hamilton function $\psi(T)$ reads:
\begin{equation}\label{GRTL GK gen L}
\dot{T}=T\cC_2-\cC_1 T,
\end{equation}
where
\be
\cC_1  =  \frac{1}{2}\,\rR_1\left(\nabla\varphi\left(T\cF^{-1}\right)\right)=
\frac{1}{2}\cF\cdot\rR\left(\cF^{-1}\nabla\varphi\left(T\cF^{-1}\right)
\right),  \label{GRTL GK gen C1}
\ee
\be
\cC_2  =  \frac{1}{2}\rR_2\left(\nabla\varphi\left(\cF^{-1}T\right)\right)=
\frac{1}{2}\rR\left(\nabla
\varphi\left(\cF^{-1}T\right)\cF^{-1}\right)\cdot\cF;
\label{GRTL GK gen C2}
\ee
this equation may be properly restricted to $\bT_{m+1}$.
The evolution of the matrices $\cT_1=I+\alpha T\cF^{-1}$ and
$\cT_2=I+\alpha\cF^{-1}T$ is described by the usual Lax equations
\begin{equation}\label{GRTL GK gen Lax}
\dot{\cT_1}=[\cT_1,\cC_1],\qquad \dot{\cT_2}=[\cT_2,\cC_2].
\end{equation}

c) For two {\rm Ad}-invariant functions $\varphi_1$, $\varphi_2$ on $\g$
the functions $\psi_1(T)=\varphi_1\left(T\cF^{-1}\right)=
\varphi_1\left(\cF^{-1}T\right)$ and
$\psi_2(T)=\varphi_2\left(T\cF^{-1}\right)=
\varphi_2\left(\cF^{-1}T\right)$ are in involution with
respect to ${\rm PB}_1(\rR,\cF)$.
\end{theorem}

{\bf Proof} follows closely the arguments used in Theorem \ref{Linear bracket 
for GTL}. We denote by $\{\cdot,\cdot\}_1$ the Poisson bracket 
${\rm PB}_1(\rR;\cF)$ on $\g$. For an arbitrary Hamilton function 
$\psi$ on $\g$, the value of the Hamiltonian vector f\/ield 
$\{\psi,\cdot\}_1$ in the point $T\in\g$ is given by:
\[
2\{\psi,T\}_1=T\rR(\nabla\psi)\cF-
\cF\rR(\nabla\psi)T+\rR^*\left(T\nabla\psi\cF-\cF\nabla\psi T\right).
\]
This may be represented in the following two alternative forms:
\begin{equation}\label{GRTL GK r proof aux2}
\{\psi,T\}_1=T\rP_{\ge 0}(\nabla\psi)\cF-\cF\rP_{\ge 0}(\nabla\psi)T
-\rP_{>0}\left(T\nabla\psi\cF-\cF\nabla\psi T\right),
\end{equation}
and
\begin{equation}\label{GRTL GK r proof aux1}
\{\psi,T\}_1=-T\rP_{<0}(\nabla\psi)\cF+\cF\rP_{<0}(\nabla\psi)T
+\rP_{\le 0}\left(T\nabla\psi\cF-\cF\nabla\psi T\right).
\end{equation}
Now take into account that $T\in\bigoplus\limits_{j=-m}^1\g_{j}$ and
$\cF\in\g_0\oplus\g_1$. Then, according to (\ref{GRTL GK r proof aux2}), 
the value of $\{\psi,T\}_1$ belongs to $\sum\limits_{j\ge -m}\g_j$. From 
the f\/irst glance, from (\ref{GRTL GK r proof aux1}) one can derive only that
$\{\psi,T\}_1$ belongs to $\sum\limits_{j\le 1}\g_j$, but, rewriting 
(\ref{GRTL GK r proof aux1}) as
\[
\alpha\{\psi,T\}_1=-\cL\rP_{<0}(\nabla\psi)\cF+\cF\rP_{<0}(\nabla\psi)\cL
+\rP_{\le 0}\left(\cL\nabla\psi\cF-\cF\nabla\psi\cL\right),
\]
we see that actually the stronger inclusion 
$\{\psi,T\}_1\in\sum\limits_{j\le 0}\g_j$
holds. Hence, the value of the vector f\/ield under consideration belongs to 
$\bigoplus\limits_{j=-m}^0\g_j$, which means that $\bT_{m+1}$ is a Poisson 
submanifold in $\g$. Other statements of the Theorem follow from the
general constructions of Sect.~\ref{Sect gen lin r-matrix}. \hfill\qed

\newpage

In the formulation and in the proof of the next Theorem we use the conventions 
(\ref{GTL convention}).
\begin{theorem}\label{linear r-matrix for GRTL of GK coord}
The bracket induced on $\bT_{m+1}$ by ${\rm PB}_1(\rR,\cF)$ has the 
following coordinate representation:
\begin{equation}\label{GRTL l br bb}
\{b_k,b_{k+1}\}_1=\alpha a_k^{(1)},
\end{equation}
\be
\left\{b_k,a_k^{(j)}\right\}_1=-a_k^{(j)},  \qquad 
\left\{a_k^{(j)},b_{k+j}\right\}_1=-a_k^{(j)}, \label{GRTL l br ba0}
\ee
\be
\left\{b_k,a_{k+1}^{(j)}\right\}_1=\alpha a_k^{(j+1)},  \qquad 
\left\{a_k^{(j)},b_{k+j+1}\right\}_1=\alpha a_k^{(j+1)}, 
\label{GRTL l br ba1}
\ee
\be
\left\{a_k^{(i)},a_{k+i}^{(j)}\right\}_1=-a_k^{(i+j)}, 
 \label{GRTL l br aa0} 
\ee
\be
\left\{a_k^{(i)},a_{k+i+1}^{(j)}\right\}_1=\alpha a_k^{(i+j+1)}. 
\label{GRTL l br aa1} 
\ee
\end{theorem}
{\bf Proof.} To calculate the induced bracket, we denote
\begin{equation}\label{GRTL GK r proof aux3}
\varphi(T)=a_k^{(i)},\qquad \psi(T)=a_{k+\ell}^{(j)},
\end{equation}
so that
\begin{equation}\label{GRTL GK r proof aux4}
\nabla\varphi=\rR(\nabla\varphi)=\lambda^i E_{k+i,k},\qquad
\nabla\psi=\rR(\nabla\psi)=\lambda^jE_{k+\ell+j,k+\ell}.
\end{equation}
We f\/ind the following value for the Poisson bracket
$\{\varphi,\psi\}_1=\left\{a_k^{(i)},a_{k+\ell}^{(j)}\right\}_1$:
\[
\ba{l}
\ds \Big\langle T,\nabla\varphi\cdot\cF\cdot\nabla\psi-
\nabla\psi\cdot\cF\cdot\nabla\varphi\Big\rangle
\vspace{3mm}\\
\ds  \qquad = \Big\langle T,\lambda^{i+j} E_{k+i,k}\cdot\Big(I
-\alpha\lambda\sum_{n=1}^N
E_{n+1,n}\Big)\cdot E_{k+\ell+j,k+\ell}\Big\rangle
\vspace{2mm}\\
\ds  \qquad -\Big\langle T,\lambda^{i+j} E_{k+\ell+j,k+\ell}\cdot
\Big(I-\alpha\lambda\sum_{n=1}^N E_{n+1,n}\Big)\cdot E_{k+i,k}\Big\rangle
\vspace{3mm}\\
\qquad  = \Big\langle T,\lambda^{i+j}(\delta_{\ell,-j}E_{k+i,k+\ell}
-\delta_{\ell,i}E_{k+\ell+j,k})
\vspace{3mm}\\
\ds \qquad -\alpha\lambda^{i+j+1}
(\delta_{\ell,-j-1}E_{k+i,k+\ell}-\delta_{\ell,i+1}E_{k+\ell+j,k})
\Big\rangle
\vspace{3mm}\\
\ds  \qquad = a_{k+\ell}^{(i+j)}\delta_{\ell,-j}-a_k^{(i+j)}\delta_{\ell,i}
-\alpha a_{k+\ell}^{(i+j+1)}\delta_{\ell,-j-1}+
\alpha a_k^{(i+j+1)}\delta_{\ell,i+1}.
\ea
\]
This proves the Theorem. \hfill \qed

\medskip

The Theorems above deliver a Hamiltonian interpretation of the f\/low 
(\ref{GRTL GK}), as well as an explanation of the Lax representations of 
this f\/low given in Proposition \ref{Lax for GRTL of GK}. Indeed, as a 
Corollary be f\/ind the following statement.

\begin{proposition} {\bf \cite{suris:K2,suris:K3}}
The flow (\ref{GRTL GK}) on 
$\cR\cT_{m+1}^{(-)}$ is Hamiltonian with respect to the linear bracket 
(\ref{GRTL l br bb})--(\ref{GRTL l br aa1}), 
with the Hamilton function
\begin{equation}\label{GRTL GK H0}
\rH_0=-\alpha^{-2}\sum_{k=1}^N \log(1+\alpha b_k).
\end{equation}
\end{proposition}

Of course, this function is singular in $\alpha$, but this can be repaired by
adding the following Casimir function:
\begin{equation}\label{GRTL GK H1}
\alpha^{-1}\rH_1=\alpha^{-1}\sum_{k=1}^N b_k+\sum_{k=1}^Na_k^{(1)};
\end{equation}
obviously, the resulting Hamilton function is regular in $\alpha$, and,
moreover, has an asymtotics
\[
\rH_0+\alpha^{-1}\rH_1=
\frac{1}{2}\sum_{k=1}^N b_k^2+\sum_{k=1}^N a_k^{(1)}+O(\alpha).
\]

The integrals of motion (\ref{GRTL GK H0}), (\ref{GRTL GK H1}) may be 
represented as the following functions of $T$:
\begin{equation}\label{GRTL GK H}
\rH_0=\varphi\left(\cF^{-1}T\right)=\varphi\left(T\cF^{-1}\right)=
-\alpha^{-2}\left(\log\det(\cT_{1,2})\right)_0, \qquad
\rH_1=\left({\rm tr}(\cT_{1,2})\right)_0.
\end{equation}
For the Hamilton function $\rH_0$ in (\ref{GRTL GK H}) we f\/ind:
\[
\cF^{-1}\nabla\varphi\left(T\cF^{-1}\right)
=\nabla\varphi\left(\cF^{-1}T\right)\cF^{-1}
=-\alpha^{-1}(\cF+\alpha T)^{-1}=-\alpha^{-1}\cL^{-1}.
\]
This explains the expressions (\ref{GRTL GK B1}), (\ref{GRTL GK B2}) for
the auxiliary matrices $\cB_{1,2}$: indeed, up to adding a matrix
proportional to $I$,
\[
\cB_1=-\alpha^{-1}\cF\cdot\pi_+\left(\cL^{-1}\right),\qquad 
\cB_2=-\alpha^{-1}\pi_+\left(\cL^{-1}\right)\cdot\cF,
\]
in agreement with (\ref{GRTL GK gen C1}), (\ref{GRTL GK gen C2}). The 
formulas (\ref{GRTL GK gen L}), (\ref{GRTL GK gen C1}), (\ref{GRTL GK gen C2}) 
def\/ine various members of the hierarchy 
${\rm RTL}_{m+1}^{(-)}(\alpha)$. The equations of motion of all 
other members, except (\ref{GRTL GK}), are much more long and complicated.
The reader should verify that this is the case already for the Hamilton 
function $\psi(T)=\varphi\left(\cF^{-1}T\right)$ such that $\nabla\varphi(T)=T$, which
corresponds to the ``f\/irst'' f\/low of the hierarchy, as opposed to the
``minus f\/irst'' f\/low ${\rm RTL}_{m+1}^{(-)}(\alpha)$. As an example, we give 
here the formulas of this ``f\/irst'' f\/low for $m=2$:
\[
\ba{l}
\ds \dot{b}_k  =  \left(1+\alpha b_k\right)\left(a_k^{(1)}-a_{k-1}^{(1)}+
  \alpha a_k^{(2)}-\alpha a_{k-2}^{(2)}\right),
\vspace{2mm}\\
\dot{a}_k^{(1)}  =  a_k^{(1)}\left(b_{k+1}-b_k+\alpha a_{k+1}^{(1)}-
  \alpha a_{k-1}^{(1)}+\alpha^2 a_{k+1}^{(2)}+\alpha^2 a_k^{(2)}-
  \alpha^2 a_{k-1}^{(2)}-\alpha^2 a_{k-2}^{(2)}\right)
\vspace{2mm}\\
\qquad  +a_k^{(2)}\left(1+\alpha b_{k+1}\right)-
  a_{k-1}^{(2)}\left(1+\alpha b_k\right),
\vspace{2mm}\\
\dot{a}_k^{(2)}  =  a_k^{(2)}\left(b_{k+2}-b_k+\alpha a_{k+2}^{(1)}-
  \alpha a_{k-1}^{(1)}+\alpha^2 a_{k+2}^{(2)}+\alpha^2 a_{k+1}^{(2)}-
  \alpha^2 a_{k-1}^{(2)}-\alpha^2 a_{k-2}^{(2)}\right).
\ea
\]

\setcounter{equation}{0}
\section{Quadratic {\mathversion{bold}$r$}-matrix 
structure for {\mathversion{bold}${\rm RTL}_{m+1}^{(-)}(\alpha)$}}
\label{Sect quadratic bracket for GRTL GK}

Now we prove a very remarkable Theorem. According to it, the quadratic 
invariant Poisson bracket of the usual Toda lattice is at the same time
invariant with respect to the f\/lows of the relativistic Toda hierarchy.

\begin{theorem}\label{Rel quadratic bracket}
Supply $\g$ with the quadratic $r$-matrix bracket 
${\rm PB}_2(\rA_1,\rA_2,\rS)$ defined by the operators (\ref{AS}). Let 
$\varphi$ be an ${\rm Ad}$-invariant function on $\g$, and consider the 
Hamiltonian flow on $\g$ with the Hamilton function
\begin{equation}\label{Rel quadr br psi}
\psi(T)=\varphi\left(T\cF^{-1}\right)=\varphi\left(\cF^{-1}T\right).
\end{equation}
The equations of motion of this flow read:
\begin{equation}\label{Rel quadr br eq}
\dot{T}=T\cC_2-\cC_1T,
\end{equation}
where
\be
\cC_1  =  \frac{1}{2}\rR_1\left(d\varphi\left(T\cF^{-1}\right)\right)=
\frac{1}{2}\cF\cdot\rR\left(\cF^{-1}d\varphi\left(T\cF^{-1}\right)\right),
\label{Rel quadr br C1}
\ee
\be
\cC_2  =  \frac{1}{2}\rR_2\left(d\varphi\left(\cF^{-1}T\right)\right)=
\frac{1}{2}\rR\left(d\varphi\left(\cF^{-1}T\right)\cF^{-1}\right)\cdot\cF.
\label{Rel quadr br C2}
\ee
\end{theorem}
{\bf Proof.} For an arbitrary Hamilton function $\psi(T)$, the Hamiltonian
equations of motion have the form (\ref{Rel quadr br eq}) with the matrices
\be
2\cC_1  =  \rA_2\left(T\nabla\psi(T)\right)+\rS^*\left(\nabla\psi(T)T\right),
\label{Rel quadr br aux1}
\ee
\be
2\cC_2  =  \rA_1\left(\nabla\psi(T)T\right)+\rS\left(T\nabla\psi(T)\right).
\label{Rel quadr br aux2}
\ee
For the functions $\psi(T)=\varphi\left(T\cF^{-1}\right)$ with ${\rm Ad}$-invariant
$\varphi$ we have, obviously:
\[
\nabla\psi(T)=\cF^{-1}\cdot\nabla\varphi\left(T\cF^{-1}\right)=\nabla\varphi\left(\cF^{-1}T\right)
\cdot\cF^{-1},
\]
so that
\[
T\nabla\psi(T)=d\varphi\left(T\cF^{-1}\right),\qquad
\nabla\psi(T)T=d\varphi\left(\cF^{-1}T\right).
\]
Hence, denoting
\[
X=\cF^{-1}\cdot d\varphi\left(T\cF^{-1}\right)=d\varphi\left(\cF^{-1}T\right)\cdot\cF^{-1},
\]
we can rewrite (\ref{Rel quadr br aux1}), (\ref{Rel quadr br aux2}) as
\be
2\cC_1  =  \rA_2(\cF X)+\rS^*(X\cF),
\label{Rel quadr br aux3}
\ee
\be
2\cC_2  =  \rA_1(X\cF)+\rS(\cF X).
\label{Rel quadr br aux4}
\ee
We now prove the following two identities. For an arbitrary matrix $X\in\g$ let
\begin{equation}
\rP_{-1}(X)=X_{-1}=\lambda^{-1}\sum_{k=1}^N x_k^{(-1)}E_{k,k+1}.
\end{equation}
Then
\be
\rA_2(\cF X)+\rS^*(X\cF)  =  \cF\rR(X)-2\alpha x_N^{(-1)}I, 
\label{Rel quadr br aux5}
\ee
\be
\rA_1(X\cF)+\rS(\cF X)  =  \rR(X)\cF-2\alpha x_N^{(-1)}I. 
\label{Rel quadr br aux6}
\ee
This will, obviously, imply the statement of the Theorem. Since 
(\ref{Rel quadr br aux5}), (\ref{Rel quadr br aux6}) are proved analogously,
we do this only for the f\/irst one of them. According to (\ref{AS}), and
$\cF=I-\alpha\cE$, we have:
\be\label{Rel quadr br aux7}
\ba{l}
\ds \rA_2(\cF X)+\rS^*(X\cF)  =  \rR_0(\cF X)+\rP_0(X\cF)-\rW(\cF X)+\rW(X\cF)
\vspace{2mm}\\
\ds \qquad =  \rR(X)-\alpha\left(\rR_0(\cE X)+\rP_0(X\cE)-\rW(\cE X)+\rW(X\cE)\right).
 \ea
\ee
The term proportional to $\alpha$ is equal to:
\[
 \ba{l}
\ds  \rP_{>0}(\cE X)-\rP_{<0}(\cE X)+\rP_0(X\cE)+\rW\left([X_{-1},\cE]\right)
\vspace{2mm}\\
\ds \qquad  =  \cE\rP_{\ge 0}(X)-\cE\rP_{<-1}(X)+X_{-1}\cE+\rW\left([X_{-1},\cE]\right)
 \vspace{2mm}\\
\ds \qquad  =  \cE\rR(X)+\cE X_{-1}+X_{-1}\cE+\rW\left([X_{-1},\cE]\right).
\ea
\]
Now a direct calculation shows that for an arbitrary matrix $X_{-1}\in\g_{-1}$
there holds:
\[
\cE X_{-1}+X_{-1}\cE+\rW\left([X_{-1},\cE]\right)=2x_N^{(-1)}I,
\]
and the proof is f\/inished. (Notice that in the open-end case the left-hand
side of the last equation simply vanishes). \hfill \qed

\medskip

The Theorem just proved yields that the matrix dif\/ferential equations
(\ref{Rel quadr br eq}) can be~proper\-ly restricted to an arbitrary
Poisson submanifold of the quadratic bracket\linebreak  ${\rm PB}_2(\rA_1,\rA_2,\rS)$,
and the resulting equations are Hamiltonian with respect to the restriction
of this bracket. In particular, this holds for the Poisson submanifold
$\bT_{m+1}$, and we f\/ind the following statement.

\begin{theorem}  The hierarchy ${\rm RTL}_{m+1}^{(-)}(\alpha)$ is Hamiltonian 
with respect to the quadratic invariant Poisson bracket (\ref{GTL q br
bb0})--(\ref{GTL q br aa3}) of the non-relativistic hierarchy 
${\rm TL}_{m+1}$.~This brac\-ket is compatible with the linear bracket of 
Theorem~\ref{linear r-matrix for GRTL of GK coord}.
\end{theorem}

In particular:

\begin{proposition}
The system (\ref{GRTL GK}) is Hamiltonian with respect to the bracket
$\{\cdot,\cdot\}_2$ on $\cR\cT_{m+1}^{(-)}$ given by the formulas
(\ref{GTL q br bb0})--(\ref{GTL q br aa3}). The corresponding
Hamilton function is equal to
\[
-\alpha\rH_0=\alpha^{-1}\sum_{k=1}^N\log(1+\alpha b_k).
\]
\end{proposition}

\newpage

\setcounter{equation}{0}
\section{Introducing the gauge transformed hierarchy}
\label{Sect GRTLs relation}

The Lax equations of ${\rm RTL}_{m+1}^{(-)}(\alpha)$ are of a somewhat
nonstandard type. We now f\/ind a~gauge transformation bringing the Lax 
equations into the standard form.
To this end, we start with the matrices $\cF(\lambda)$, $\cL(a,b,\lambda)$
for the ${\rm RTL}_{m+1}^{(-)}(-\alpha)$ hierarchy, i.e. we change $\alpha$
to $-\alpha$ in the formulas (\ref{GRTL GK F}), (\ref{GRTL GK L}). As it 
follows from Theorems \ref{linear r-matrix for GRTL of GK}, \ref{Rel 
quadratic bracket}, we have the following Lax representations for the f\/lows 
of the ${\rm RTL}_{m+1}^{(-)}(-\alpha)$ hierarchy:
\begin{equation}\label{GRTL relation aux1}
\dot{\cT}_i=[\cT_i,\cB_i]=[\cA_i,\cT_i],\qquad i=1,2,
\end{equation}
where
\be
\cB_1=\cF\cdot\pi_+\left(\cF^{-1}f\left(\cL\cF^{-1}\right)\right), \qquad 
\cA_1=\cF\cdot\pi_-\left(\cF^{-1}f\left(\cL\cF^{-1}\right)\right), 
\label{GRTL relation aux2}
\ee
\be
\cB_2=\pi_+\left(f\left(\cF^{-1}\cL\right)\cF^{-1}\right)\cdot \cF, \qquad 
\cA_2=\pi_-\left(f\left(\cF^{-1}\cL\right)\cF^{-1}\right)\cdot\cF, 
\label{GRTL relation aux3}
\ee
with an Ad-covariant function $f:\g\mapsto\g$ (we have 
$f(\cT)=\nabla\varphi((\cT-I)/\alpha)$ and $f(\cT)=d\varphi((\cT-I)/\alpha)$ 
for the Hamiltonian f\/lows with the Hamilton function 
$\varphi\!\left(T\cF^{-1}\right)=\varphi\!\left(\cF^{-1}T\right)$ in the linear and the quadratic 
bracket, respectively).

Now introduce two arbitrary diagonal matrices
\begin{equation}\label{GRTL Omegas}
\Omega_1={\rm diag}(\omega_1,\omega_2,\ldots,\omega_N),\qquad
\Omega_2={\rm diag}(\omega_1',\omega_2',\ldots,\omega_N').
\end{equation}
It will be convenient for us to denote
\begin{equation}\label{GRTL ab to cd Lax}
\Omega_i\cdot\cT_i\cdot\Omega_i^{-1}=T_i^{-1},\qquad i=1,2.
\end{equation}
The gauge transformed Lax equations for the matrices $T_i$ are easily derived
from (\ref{GRTL relation aux1}) and (\ref{GRTL ab to cd Lax}):
\begin{equation}\label{GRTL relation aux4}
\dot{T}_i=[T_i,B_i]=[A_i,T_i],\qquad i=1,2,
\end{equation}
where
\begin{equation}\label{GRTL relation aux5}
B_i=\Omega_i\cB_i\Omega_i^{-1}-\dot{\Omega}_i\Omega_i^{-1},\qquad
A_i=\Omega_i\cA_i\Omega_i^{-1}+\dot{\Omega}_i\Omega_i^{-1}.
\end{equation}

>From the expressions (\ref{GRTL relation aux2}), 
(\ref{GRTL relation aux3}) one sees immediately that
\[
\cB_i\in\g_{\ge 0}, \qquad \cA_i\in\g_{\le 0},
\]
and the same holds for the matrices $B_i$, $A_i$:
\begin{equation}\label{GRTL relation aux7}
B_i\in\g_{\ge 0}, \qquad A_i\in\g_{\le 0}.
\end{equation}
Our aim is to assure that 
\begin{equation}\label{GRTL relation want}
B_i\in\g_{\ge 0}=\g_+, \qquad A_i\in\g_{<0}=\g_-.
\end{equation}
To this end it is necessary only to kill the diagonal entries of $A_i$, 
i.e. to def\/ine the entries of the matrices $\Omega_i$ in such a way that
\begin{equation}\label{GRTL relation want2}
\dot{\Omega}_i\Omega_i^{-1}=-\rP_0(\cA_i).
\end{equation}
So, all we need to do, is to determine the diagonal parts of the matrices
$\cA_i$.
\begin{lemma}\label{Lemma for the relativistic gauge}
If
\[
\rP_0(\cA_1)={\rm diag}(\ga_1,\ldots,\ga_N),\qquad
\rP_0(\cA_2)={\rm diag}(\ga_1',\ldots,\ga_N'),
\]
then
\[
\ga_k=C+\sum_{j=1}^{k-1}\frac{\alpha\dot{b}_j}{1-\alpha b_j},\qquad
\ga_k'=C+\sum_{j=1}^{k}\frac{\alpha\dot{b}_j}{1-\alpha b_j}\;,
\]
where $C$ is some constant (independent on $k$).
\end{lemma}
{\bf Proof.} Consider the matrix equations
\begin{equation}\label{GRTL relation aux9}
\cA_1\cF-\cF\cA_2=0, \qquad \dot{\cL}=\cA_1\cL-\cL\cA_2.
\end{equation}
Due to $\cA_i\in\g_{\le 0}$ and the def\/inition of $\cF$, we f\/ind from the
f\/irst equation in (\ref{GRTL relation aux9}):
\begin{equation}\label{GRTL relation aux10}
\ga_k'=\ga_{k+1}.
\end{equation}
On the other hand, since $\cL\in\g_{\le 0}$, we f\/ind, considering the $\g_0$
part of the second equation in (\ref{GRTL relation aux9}):
\[
-\alpha\dot{b}_k=(1-\alpha b_k)(\ga_k-\ga_k')\quad\Rightarrow\quad
\ga_{k+1}-\ga_k=\frac{\alpha\dot{b}_k}{1-\alpha b_k}.
\]
This implies the statement of the Lemma. \hfill \qed

\medskip

So, let us def\/ine the entries of the diagonal matrices $\Omega_i$, $i=1,2$, by 
the following formulas:
\begin{equation}\label{GRTL omegas}
\omega_k=\delta^{k-1}\prod_{j=1}^{k-1}(1-\alpha b_j), \qquad
\omega_k'=\delta^{k-1}\prod_{j=1}^k(1-\alpha b_j),
\end{equation}
where we set in the periodic case 
\[
\delta^N=\prod_{j=1}^N (1-\alpha b_j)^{-1},
\]
and in the open-end case simply $\delta=1$. Of course, in the periodic case
$\delta$ is a spectral invariant of the Lax matrix, and therefore an integral 
of motion of an arbitrary f\/low of the hierarchy. 

\begin{theorem}\label{GRTLs relation Lax}
Fix an ${\rm Ad}$-covariant function $f:\g\mapsto\g$ and consider the 
corresponding flow of the hierarchy ${\rm RTL}_{m+1}^{(-)}(-\alpha)$. Then 
the evolution of the matrices $T_i$ is 
described by the standard Lax equations
\[
\dot{T}_i= \left[T_i,\pi_+\left(f\left(T_i^{-1}\right)\right)\right]=
\left[\pi_-\left(f\left(T_i^{-1}\right)\right),T_i\right], \qquad i=1,2.
\]
\end{theorem}
{\bf Proof.} From Lemma \ref{Lemma for the relativistic gauge} and the
def\/initions (\ref{GRTL omegas}) we f\/ind:
\[
\ga_k=C-\frac{\dot{\omega}_k}{\omega_k},\qquad 
\ga_k'=C-\frac{\dot{\omega}_k'}{\omega_k'},
\]
which means that
\begin{equation}\label{GRTL relation aux8}
\rP_0(\cA_i)+\dot{\Omega}_i\Omega_i^{-1}=C\cdot I.
\end{equation}
Therefore, renaming $A_i-C\cdot I$ by $A_i$, and $B_i+C\cdot I$ by $B_i$ 
(which does not af\/fect the Lax equations (\ref{GRTL relation aux4})), we come 
to the desired property (\ref{GRTL relation want}). Further, from 
(\ref{GRTL relation aux2}), (\ref{GRTL relation aux3}) we derive:
\[
\cB_i+\cA_i=f(\cT_i)
\]
and it follows that
\begin{equation}\label{GRTL relation aux6}
B_i+A_i=f\left(T_i^{-1}\right).
\end{equation}
Therefore $B_i$ and $A_i$ are with necessity the $\pi_+$ and $\pi_-$ 
projections of $f(T_i^{-1})$, respectively.~\hfill\qed

\medskip

In particular, the f\/low ${\rm RTL}_{m+1}^{(-)}(-\alpha)$ is characterized, as 
we have seen, by $f(\cT)=\alpha^{-1}\cT^{-1}$. Therefore its gauge transformed Lax representation
reads:
\[
\dot{T}_i=\alpha^{-1}[T_i,\pi_+(T_i)].
\]
In other words, the image of the f\/low ${\rm RTL}_{m+1}^{(-)}(-\alpha)$ in the
gauge transformed hierarchy becomes the ``f\/irst'' f\/low.

\setcounter{equation}{0}
\section[Multi-f\/ield RTL: second version]
{Multi-f\/ield analog of the relativistic \\ Toda lattice: the second 
construction}
\label{Sect GRTL}

An easy calculation shows that (we still consider $\cL$, $\cF$ attached to
the ${\rm RTL}_{m+1}^{(-)}(-\alpha)$ hierarchy):
\[
\Omega_1\cdot\cF(\lambda)\cdot\Omega_2^{-1}=
L\left(d,c^{(1)},\ldots,c^{(m)},\delta\lambda\right),
\]
\[
\Omega_1\cdot\cL\left(b,a^{(1)},\ldots,a^{(m)},\lambda\right)\cdot\Omega_2^{-1}=
U\left(d,c^{(1)},\ldots,c^{(m)},\delta\lambda\right),
\]
where 
\be
L\left(d,c^{(1)},\ldots,c^{(m)},\lambda\right)  = 
\sum_{k=1}^N (1+\alpha d_k)E_{kk}+\alpha\lambda\sum_{k=1}^NE_{k+1,k},
\label{GRTL L}
\ee
\be
U\left(d,c^{(1)},\ldots,c^{(m)},\lambda\right)  =  I-\alpha\sum_{j=1}^m\lambda^{-j}
\sum_{k=1}^N c_k^{(j)}E_{k,k+j},
\label{GRTL U}
\ee
and the variables $d$, $c^{(j)}$ are def\/ined by the formulas 
\begin{equation}\label{GRTLs ab to cd}
\bB(\alpha):\quad  d_k=\frac{b_k}{1-\alpha b_k},\qquad 
c_k^{(j)}=\frac{a_k^{(j)}}{\prod\limits_{i=0}^j(1-\alpha b_{k+i})}.
\end{equation}

We want now to study the gauge transformed hierarchy in its own rights.
The phase space of this second version of the generalized relativistic Toda 
lattice with $m+1$ f\/ields, abbreviated ${\rm RTL}_{m+1}^{(+)}(\alpha)$, is, 
in the periodic case, the space
\begin{equation}\label{GRTL phase sp}
\cR\cT_{m+1}^{(+)}={\mathbb R}^{(m+1)N}\left(d,c^{(1)},\ldots,c^{(m)}\right).
\end{equation}
Here
\[
d=(d_1,\ldots,d_N) \quad {\rm and}\quad 
c^{(j)}=\left(c_1^{(j)},\ldots,c_N^{(j)}\right) \quad (j=1,\ldots,m)
\]
are the $m+1$ f\/ields. The Lax matrix map 
$(L,U^{-1}):\cR\cT_{m+1}^{(+)}\mapsto\g\otimes\g$ is def\/ined by the 
formulas (\ref{GRTL L}), (\ref{GRTL U}). We set also
\begin{equation}\label{GRTL Ts}
T_1=LU^{-1},\qquad T_2=U^{-1}L.
\end{equation}

\begin{proposition} \label{Lax for GRTL}
The Lax triads
\begin{equation}\label{GRTL Lax triads}
\dot{L}=LB_2-B_1L,\qquad \dot{U}=UB_2-B_1U
\end{equation}
with the Lax matrices (\ref{GRTL L}), (\ref{GRTL U}) and the auxiliary 
matrices
\be
B_1\left(d,c^{(1)},\ldots,c^{(m)},\lambda\right)  =  
\sum_{k=1}^N \left(d_k+\alpha c_{k-1}^{(1)}\right)E_{kk}+\lambda\sum_{k=1}^NE_{k+1,k},
\label{GRTL A1}
\ee
\be
B_2\left(d,c^{(1)},\ldots,c^{(m)},\lambda\right)  = 
\sum_{k=1}^N \left(d_k+\alpha c_k^{(1)}\right)E_{kk}+\lambda\sum_{k=1}^NE_{k+1,k},
\label{GRTL A2} 
\ee
are equivalent to the following system of differential equations:
\begin{equation}\label{GRTL}
\left\{\begin{array}{l}
\dot{d}_k=(1+\alpha d_k)\left(c_k^{(1)}-c_{k-1}^{(1)}\right), 
\vspace{2mm}\\
\dot{c}_k^{(j)}=c_k^{(j)}\left(d_{k+j}-d_k+
\alpha c_{k+j}^{(1)}-\alpha c_{k-1}^{(1)}\right)+
\left(c_k^{(j+1)}-c_{k-1}^{(j+1)}\right), 
\vspace{2mm}\\
\dot{c}_k^{(m)}=c_k^{(m)}\left(d_{k+m}-d_k+\alpha c_{k+m}^{(1)}-
\alpha c_{k-1}^{(1)}\right), \quad 1\le j\le m-1.
\end{array}\right.
\end{equation}
\end{proposition}
{\bf Proof} -- an elementary matrix calculation.\hfill \qed

\medskip

Notice that
\begin{equation}\label{GRTL As}
B_1=\pi_+\left(\left(LU^{-1}-I\right)/\alpha\right),\qquad
B_2=\pi_+\left(\left(U^{-1}L-I\right)/\alpha\right).
\end{equation}
So, the above Lax representation is indeed of a standard type.

One should compare the equations of motion (\ref{GRTL}) of 
${\rm RTL}_{m+1}^{(+)}(\alpha)$ with the
``f\/irst'' f\/low of the ${\rm RTL}_{m+1}^{(-)}(\alpha)$ hierarchy 
(given for $m=2$ at the end of Sect.~\ref{Sect r-matrix for GRTL of GK}).

Returning to the relation between the hierarchies
${\rm RTL}_{m+1}^{(+)}(\alpha)$ and ${\rm RTL}_{m+1}^{(-)}(-\alpha)$, we 
see that it is established via the change of variables
\[
\bB(\alpha):\quad\cR\cT_{m+1}^{(-)}\left(b,a^{(1)},\ldots,a^{(m)}\right)\mapsto
\cR\cT_{m+1}^{(+)}\left(d,c^{(1)},\ldots,c^{(m)}\right)
\]
given by the formulas (\ref{GRTLs ab to cd}). Clearly, this transformation
is invertible, and its inverse is given by
\begin{equation}\label{GRTLs cd to ab}
\bB^{-1}(\alpha)=\bB(-\alpha):\quad b_k=\frac{d_k}{1+\alpha d_k},\qquad 
a_k^{(j)}=\frac{c_k^{(j)}}{\prod\limits_{i=0}^j(1+\alpha d_{k+i})}.
\end{equation}
This transformation relates, as it follows from Theorem 
\ref{GRTLs relation Lax}, the whole hierarchies. In particular, this holds
for the simplest f\/lows: the change of variables given by the formulas
(\ref{GRTLs ab to cd}) brings the system ${\rm RTL}_{m+1}^{(-)}(-\alpha)$
(\ref{GRTL GK}) into the system ${\rm RTL}_{m+1}^{(+)}(\alpha)$ (\ref{GRTL}),
and vice versa.

\setcounter{equation}{0}
\section[Quadratic $r$-matrix structure for ${\rm RTL}_{m+1}(\alpha)$]
{Quadratic {\mathversion{bold}$r$}-matrix structure for 
{\mathversion{bold}${\rm RTL}_{m+1}^{(+)}(\alpha)$}}
\label{Sect qudratic bracket for GRTL}

Consider the subset
\[
\bR\bT_{m+1}=\left(\cE\oplus\g_0\right)\times
\left(I\oplus\bigoplus_{j=1}^m\g_{-j}\right)^{-1}
\]
of $\g\otimes\g$, consisting of pairs $(L,U^{-1})$ of matrices
(\ref{GRTL L}), (\ref{GRTL U}). We show now that this subset is a Poisson 
submanifold, if $\g\otimes\g$ is equipped with a certain quadratic 
$r$-matrix bracket. Actually, this bracket was used already in \cite{suris:S1,suris:S2}
to give an $r$-matrix interpretation of the usual (two-f\/ield) relativistic
Toda lattice and of the Volterra lattice.

\begin{theorem} \label{r-matrix for GRTL in g+g}
a) Supply the algebra $\g\otimes\g$ with the bracket 
$\alpha\,\!{\rm PB}_2(\bA_1,\bA_2,\bS)$ defined by the operators
\begin{equation}\label{VL Ops A in g+g}
\bA_1=\left(\begin{array}{cc} \rA_1 & -\rS\\ \rS^* & \rA_1\end{array}\right),
\qquad
\bA_2=\left(\begin{array}{cc} \rA_2 & -\rS^*\\ \rS & \rA_2\end{array}\right),
\end{equation}
\begin{equation}\label{VL Ops S in g+g}
\bS=\left(\begin{array}{cc} \rS & \rS \\ \rS & -\rS^* \end{array}\right),
\qquad
\bS^*=\left(\begin{array}{cc} \rS^* & \rS^*\\ \rS^* & -\rS\end{array}\right),
\end{equation}
where the operators $\rA_1$, $\rA_2$, $\rS$, $\rS^*$ are as in 
(\ref{AS}). Then the set $\bR\bT_{m+1}$ is a Poisson submanifold
in $\g\otimes\g$.

b) Consider the monodromy maps $\bM_{1,2}:\g\otimes\g\mapsto\g$
\begin{equation}\label{GRTL monodromy}
\bM_1:(L,U^{-1})\mapsto T_1=LU^{-1},
\qquad \bM_2: (L,U^{-1})\mapsto T_2=U^{-1}L.
\end{equation}
Both maps $\bM_{1,2}$ are Poisson, if the target space $\g$ is equipped 
with the Poisson bracket $\alpha\,\!{\rm PB}_2(\rA_1,\rA_2,\rS)$. 

c) Let $\varphi$ be an ${\rm Ad}$-invariant function
on $\g$. Then the Hamiltonian equations of motion on $\g\otimes\g$
with the Hamilton function $\varphi\circ\bM_{1,2}$ may be presented in the 
form of the ``Lax triads''
\begin{equation}\label{GRTL hier Lax triads}
\dot{L}=LC_2-C_1L,\qquad \dot{U}=UC_2-C_1U,
\end{equation}
where
\begin{equation}\label{GRTL Cs}
C_1=\frac{1}{2}\rR\left(d\varphi\left(LU^{-1}\right)\right),\qquad
C_2=\frac{1}{2}\rR\left(d\varphi\left(U^{-1}L\right)\right).
\end{equation} 
\end{theorem}
{\bf Proof.} As usual, we have to prove that an arbitrary Hamiltonian
vector f\/ield on 
\[
\left(\g\otimes\g,{\rm PB}_2(\bA_1,\bA_2,\bS)\right)
\] 
is tangent to $\bR\bT_{m+1}$. It is enough to consider separately vector
f\/ields of Hamilton
functions $\phi(L)$ and $\varphi(U)$, depending only on the f\/irst, resp. 
second, factor in the tensor product, because an arbitrary Hamiltonian vector
f\/ield is a linear combination of such ones. Further, the calculations are
simplif\/ied, if we regard the values of these vector f\/ields, parametrizing the 
second factors via $U$ instead of $U^{-1}$. With these conventions,
the value of the vector f\/ield of the Hamilton function $\phi(L)$  
in the point of $\g\otimes\g$ parametrized as $(L,U)$, is given~by
\be
2\alpha^{-1}\{\phi(L),L\}_{2\alpha} = 
L\cdot \rA_1(\nabla\phi L)-
  \rA_2(L\nabla\phi)\cdot L+L\cdot\rS(L\nabla\phi)-\rS^*(\nabla\phi L)\cdot L,
 \label{GRTL vf1}
\ee
\be
2\alpha^{-1}\{\phi(L),U\}_{2\alpha} = 
U\cdot \rS^*(\nabla\phi L)-
  \rS(L\nabla\phi)\cdot U+U\cdot\rS(L\nabla\phi)-\rS^*(\nabla\phi L)\cdot U.
  \qquad\label{GRTL vf2}
\ee 
Similarly, for the vector f\/ield of the Hamilton function $\varphi(U)$  
in the point of $\g\otimes\g$ parametrized as $(L,U)$, is given by
\be
2\alpha^{-1}\{\varphi(U),L\}_{2\alpha} = 
  -L\cdot \rS(\nabla\varphi U)+
  \rS^*(U\nabla\varphi)\cdot L+L\cdot\rS(U\nabla\varphi)-
  \rS^*(\nabla\varphi U)\cdot L,
\label{GRTL vf3}
\ee
\be
2\alpha^{-1}\{\varphi(U),U\}_{2\alpha} =
-U\cdot \rA_2(\nabla\varphi U)+
  \rA_1(U\nabla\varphi)\cdot U+U\cdot\rS(U\nabla\varphi)-
  \rS^*(\nabla\varphi U)\cdot U.
  \label{GRTL vf4}
\ee

Consider the f\/irst vector f\/ield $\{\phi(L),\cdot\}_{2\alpha}$. Its 
$L$-component (\ref{GRTL vf1}) lies in $\g_0$: this is a~particular case of 
Theorem \ref{Quadratic bracket for GTL}. The value of its $U$-component
(\ref{GRTL vf2}) lies, obviously, in $\sum\limits_{j=0}^m\g_{-j}$, because the
range of the operators $\rS$, $\rS^*$ is $\g_0$. Moreover, it is easy to
see that the $\g_0$-component of (\ref{GRTL vf2}) vanishes. Henceforth,
the vector f\/ields $\{\phi(L),\cdot\}_{2\alpha}$ are tangent to~$\bR\bT_{m+1}$.

Let us turn now to the second vector f\/ield $\{\varphi(U),\cdot\}_{2\alpha}$.
Its $L$-component, (\ref{GRTL vf3}), may be rewritten as
\begin{equation}\label{GRTL proof aux}
\ba{l}
2\alpha^{-1}\{\varphi(U),L\}_{2\alpha} =
  LD+DL+\rW(D)L-L\rW(D)
\vspace{2mm}\\
\ds \qquad -LD'-D'L-\rW(D')L+L\rW(D'),
\ea
\end{equation}
where
\[
D=\rP_0(U\nabla\varphi),\qquad D'=\rP_0(\nabla\varphi U).
\]
The vanishing of the $\g_1$-component of this expression follows from 
(\ref{TL proof2 aux6}). Finally, for the $U$-component (\ref{GRTL vf4})
of the vector f\/ield under consideration, we have the following two 
equivalent expressions:
\be
\ba{l}
2\alpha^{-1}\{\varphi,U\}_{2\alpha} = 
-2U\rP_{>0}(\nabla\varphi U)+2\rP_{>0}(U\nabla\varphi)U   
\vspace{2mm}\\
\ds \qquad -U\rP_0(\nabla\varphi U)+\rP_0(U\nabla\varphi)U+
U\rP_0(U\nabla\varphi)-\rP_0(\nabla\varphi U)U   
\vspace{2mm}\\
\ds \qquad +U\rW(\nabla\varphi U)+\rW(U\nabla\varphi)U
-U\rW(U\nabla\varphi)-\rW(\nabla\varphi U)U,
\ea \label{GRTL proof vf1}
\ee
and
\be
\ba{l}
2\alpha^{-1}\{\varphi,U\}_{2\alpha}
= 2U\rP_{<0}(\nabla\varphi U)-2\rP_{<0}(U\nabla\varphi)U   
\vspace{2mm}\\
\ds \qquad +U\rP_0(\nabla\varphi U)-\rP_0(U\nabla\varphi)U+
U\rP_0(U\nabla\varphi)-\rP_0(\nabla\varphi U)U            
\vspace{2mm}\\
\ds \qquad +U\rW(\nabla\varphi U)+\rW(U\nabla\varphi)U
-U\rW(U\nabla\varphi)-\rW(\nabla\varphi U)U.
\ea \label{GRTL proof vf2}
\ee
(as compared with the formulas (\ref{GTL proof2 vf1}) and 
(\ref{GTL proof2 vf2}), respectively, in our present formulas the overall 
signs in the f\/irst lines and in the f\/irst two terms on the second lines 
are opposite). The f\/irst expression above assures that for 
$U\in I\oplus\bigoplus\limits_{j=-m}^{-1}\g_{j}$ this vector belongs to 
$\bigoplus\limits_{j\ge -m}\g_j$. The second expression yields that this vector 
belongs also to $\bigoplus\limits_{j\le 0}\g_j$; moreover, since the 
$\g_0$ component of $U$ is equal to $I$, it is easy to see that the 
$\g_0$ component of (\ref{GRTL proof vf2}) vanishes, so that actually it 
belongs to $\bigoplus\limits_{j\le -1}\g_j$. Hence, it belongs to 
$\bigoplus\limits_{j=-m}^{-1}\g_j$.  This f\/inishes the proof of the f\/irst
statement of the Theorem. The parts b) and c) follow from Theorem 
\ref{monodromy} with $n=2$. \hfill  \qed

\begin{theorem} \label{r-matrix for GRTL in g+g coord}
The coordinate representation of the bracket induced by
$\alpha{\rm PB}_2(\bA_1,\bA_2,\bS)$ on the submanifold $\bR\bT_{m+1}$,
is given by the following formulas: (in (\ref{GRTL br cc1}), 
(\ref{GRTL br cc2}) below we assume that $i\le j$):
\be
\left\{d_k,c_k^{(j)}\right\}_{2\alpha}  =  -c_k^{(j)}(1+\alpha d_k),  \qquad  
\left\{c_k^{(j)},d_{k+j}\right\}_{2\alpha}=-c_k^{(j)}(1+\alpha d_{k+j}), 
 \label{GRTL br dc}
\ee
\be
\left\{c_k^{(i)},c_{k+i}^{(j)}\right\}_{2\alpha}  =  
-c_k^{(i+j)}-\alpha c_k^{(i)}c_{k+i}^{(j)},
\label{GRTL br cc0}
\ee
\be
\left\{c_k^{(i)},c_{k+\ell}^{(j)}\right\}_{2\alpha}  = 
-\alpha c_k^{(i)}c_{k+\ell}^{(j)}
+\alpha c_k^{(j+\ell)}c_{k+\ell}^{(i-\ell)} \qquad (1\le\ell\le i-1),   
\label{GRTL br cc1}
\ee
\be
\left\{c_k^{(j)},c_{k+\ell}^{(i)}\right\}_{2\alpha}  = 
-\alpha c_k^{(j)}c_{k+\ell}^{(i)}
+\alpha c_k^{(i+\ell)}c_{k+\ell}^{(j-\ell)} \qquad (j-i+1\le\ell\le j-1). 
\label{GRTL br cc2}
\ee
\end{theorem}
{\bf Proof.} The vanishing of all Poisson brackets $\{d_k,d_\ell\}_{2\alpha}$
is a light consequence of (\ref{GRTL vf1}). Similarly, (\ref{GRTL br dc}) 
follows simply by considering the $\g_0$ component of (\ref{GRTL proof aux}).
The longest is the calculation of Poisson brackets among the $c_k^{(j)}$
variables. To this end we use the def\/ining formula, following from
(\ref{GRTL proof vf1}):
\begin{eqnarray}\label{GRTL proof PB}
\lefteqn{2\alpha^{-1}\{\varphi(U),\psi(U)\}_{2\alpha} =
-2\langle\rP_{>0}(\nabla\varphi U),\nabla\psi U\rangle
+2\langle\rP_{>0}(U\nabla\varphi),U\nabla\psi\rangle }\nonumber\\ 
&& \hspace*{-10pt}-\langle\rP_0(\nabla\varphi U),\nabla\psi U\rangle+
\langle\rP_0(U\nabla\varphi),U\nabla\psi\rangle+        
\langle\rP_0(U\nabla\varphi),\nabla\psi U\rangle
-\langle\rP_0(\nabla\varphi U),U\nabla\psi\rangle\nonumber\\
&&  \hspace*{-10pt}+\langle\rW(\nabla\varphi U),\nabla\psi U\rangle+
\langle\rW(U\nabla\varphi),U\nabla\psi\rangle
 -\langle \rW(U\nabla\varphi),\nabla\psi U\rangle-
\langle\rW(\nabla\varphi U),U\nabla\psi\rangle.\nonumber\\
\end{eqnarray}
Set in this formula
\[
\varphi(U)=-\alpha c_k^{(i)},\qquad \psi(U)=-\alpha c_{k+\ell}^{(j)},
\]
so that
\[
\nabla\varphi=\lambda^iE_{k+i,i},\qquad
\nabla\psi=\lambda^jE_{k+\ell+j,k+\ell}.
\]
We f\/ind:
\begin{equation}\label{GRTL proof aux1}
\nabla\varphi U=\lambda^iE_{k+i,k}
-\alpha
\sum_{\beta=1}^m\lambda^{i-\beta}c_k^{(\beta)}E_{k+i,k+\beta},
\end{equation}
\begin{equation}\label{GRTL proof aux2}
U\nabla\varphi=\lambda^iE_{k+i,k}-\alpha
\sum_{\beta=1}^m\lambda^{i-\beta}c_{k+i-\beta}^{(\beta)}E_{k+i-\beta,k}.
\end{equation}
We consider f\/irst the contribution to the Poisson bracket 
$\alpha^{-1}\{\varphi,\psi\}_{2\alpha}=
\alpha\left\{c_k^{(i)},c_{k+\ell}^{(j)}\right\}_{2\alpha}$ 
from the f\/irst line in (\ref{GRTL proof PB}):
\[
\ba{l}
\ds 
\left\langle \lambda^iE_{k+i,k}-\alpha
\sum_{\beta=1}^{i-1}\lambda^{i-\beta}c_k^{(\beta)}E_{k+i,k+\beta},
\alpha\sum_{\gamma=j+1}^m\lambda^{j-\gamma}c_{k+\ell}^{(\gamma)}
E_{k+\ell+j,k+\ell+\gamma}\right\rangle \qquad\qquad\qquad
\vspace{3mm}\\
\ds \quad -\left\langle \lambda^iE_{k+i,k}-\alpha
\sum_{\beta=1}^{i-1}\lambda^{i-\beta}
c_{k+i-\beta}^{(\beta)}E_{k+i-\beta,k},\alpha
\sum_{\gamma=j+1}^m\lambda^{j-\gamma}c_{k+\ell+j-\gamma}^{(\gamma)}
E_{k+\ell+j-\gamma,k+\ell}\right\rangle. 
\ea
\]
Calculating these scalar products, we f\/ind:
\[
\ba{l}
\ds  =\alpha\sum_{\gamma=j+1}^m\left(
c_{k+\ell}^{(\gamma)}\,\delta_{0,\ell+j}\delta_{i,\ell+\gamma}-
c_{k+\ell+j-\gamma}^{(\gamma)}\delta_{0,\ell+j-\gamma}
\delta_{i,\ell}\right)  
\vspace{3mm}\\
\ds \qquad  -\alpha^2\sum_{\beta=1}^{i-1}\sum_{\gamma=j+1}^m\left(
c_k^{(\beta)}c_{k+\ell}^{(\gamma)}\delta_{\beta,\ell+j}
\delta_{i,\ell+\gamma}
-c_{k+i-\beta}^{(\beta)}c_{k+\ell+j-\gamma}^{(\gamma)}
\delta_{0,\ell+j-\gamma}\delta_{i-\beta,\ell}\right)
\vspace{3mm}\\
\ds  \qquad  =\alpha c_{k+\ell}^{(i+j)}\delta_{\ell,-j}-\alpha
c_k^{(i+j)}\delta_{\ell,i}-\alpha^2\bar{\varkappa}_{k\ell}^{(ij)}
c_k^{(\ell+j)}c_{k+\ell}^{(i-\ell)},
\ea
\]
where
\be\label{barkappa}
\ba{l}
\bar{\varkappa}_{k\ell}^{(ij)}  =   
\chi_{\ell}(-j+1,i-j-1)\chi_{\ell}(i-m,i-j-1)
-\chi_{\ell}(1,m-j)\chi_{\ell}(1,i-1)\vspace{2mm}\\ 
\ds \qquad  =    \chi_{\ell}\left(-\min (j-1,m-i),i-j-1\right)
-\chi_{\ell}\left(1,\min(i-1,m-j)\right).
\ea
\ee
Assuming, for the sake of def\/initeness, that $i\le j$, we see that the 
intervals of the two characteristic functions in the last line do not
intersect. So, we found the contributions to the Poisson bracket
$\alpha\left\{c_k^{(i)},c_{k+\ell}^{(j)}\right\}_{2\alpha}$ described by the 
f\/irst term on the right-hand side of the formula (\ref{GRTL br cc0}), and the 
second terms on the right-hand sides of (\ref{GRTL br cc1}), 
(\ref{GRTL br cc2}). 

Finally, calculating the contribution of the remaining part of 
(\ref{GRTL proof PB}), we get:
\begin{equation}\label{GRTL br aux3}
\alpha^2\bar{\varepsilon}_{k\ell}^{(ij)}c_k^{(i)}c_{k+\ell}^{(j)},
\end{equation}
where
\begin{equation}\label{GRTL br aux4}
\ba{l}
\bar{\varepsilon}_{k\ell}^{(ij)}=\frac{1}{2}(
-\delta_{i,\ell+j}+\delta_{0,\ell}+\delta_{0,\ell+j}-\delta_{i,\ell}
\vspace{2mm}\\
\qquad +w_{k+i,k+\ell+j}+w_{k,k+\ell}-w_{k,k+\ell+j}-w_{k+i,k+\ell}).
\ea
\end{equation}
This coef\/f\/icient dif\/fers from (\ref{GTL q br aux3})
by $-\delta_{i,\ell+j}+\delta_{0,\ell}$, so that the result (for $i\le j$)
may be written down immediately:
\begin{equation}\label{GRTL br aux5}
\bar{\varepsilon}_{k\ell}^{(ij)}=
\chi_{\ell}(-j,-j+i-1)-\chi_{\ell}(1,i).
\end{equation}
This describes the f\/irst terms on the right-hand sides of 
(\ref{GRTL br cc1}), (\ref{GRTL br cc2}), which f\/inishes
the proof of the part a) of Theorem \ref{r-matrix for GRTL in g+g} and the 
proof of Theorem \ref{r-matrix for GRTL 
in g+g coord}. The parts b) and c) of Theorem \ref{r-matrix for GRTL in g+g}
follow from Theorem \ref{monodromy} with $m=2$. \hfill \qed

\medskip

\noindent
{\bf Corollary.} {\it The system ${\rm RTL}_{m+1}^{(+)}(\alpha)$ 
(\ref{GRTL}) is Hamiltonian with respect to the bracket
(\ref{GRTL br dc})--(\ref{GRTL br cc2}), with the Hamilton function
\begin{equation}\label{GRTL H}
\alpha^{-1}\rH_1=\alpha^{-1}\sum_{k=1}^N d_k+\sum_{k=1}^N c_k^{(1)}=
\alpha^{-1}\left({\rm tr}(LU^{-1}-I)\right)_0.
\end{equation}
}

Unfortunately, the ${\rm RTL}_{m+1}^{(+)}(\alpha)$ hierarchy seems not 
to have an invariant linear Poisson bracket. Nevertheless, it is
bi-Hamiltonian, and the most direct way to f\/ind its second Hamiltonian
formulation is to push forward under $\bB(\alpha)$ the known invariant
Poisson brackets of the ${\rm RTL}_{m+1}^{(-)}(-\alpha)$ hierarchy. 
It can be verif\/ied that the bracket $\{\cdot,\cdot\}_{2\alpha}$ found in 
Theorem \ref{r-matrix for GRTL in g+g coord} is recovered by applying this
construction to the bracket $\{\cdot,\cdot\}_1-\alpha\{\cdot,\cdot\}_2$
of the hierarchy ${\rm RTL}_{m+1}^{(-)}(-\alpha)$. A compatible bracket 
appears, if we start just with $\{\cdot,\cdot\}_1$. We formulate this as 
a separate statement.

\begin{theorem}\label{GRTL cubic bracket}
The hierarchy ${\rm RTL}_{m+1}^{(+)}(\alpha)$ has 
an invariant cubic Poisson bracket $\{\cdot,\cdot\}_{3\alpha}$, which can be 
defined as the push-forward under the change of variables $\bB(\alpha)$
(\ref{GRTLs ab to cd}) of the invariant bracket $\{\cdot,\cdot\}_1$ of 
the ${\rm RTL}_{m+1}^{(-)}(-\alpha)$ hierarchy. The brackets 
$\{\cdot,\cdot\}_{2\alpha}$ and $\{\cdot,\cdot\}_{3\alpha}$ are compatible.
\end{theorem}

We shall not try to write down the general formulas for this pushed
bracket, as they turn out to be complicated enough. However, in the next
Section we give the result of calculations for this bracket in the particular 
case~$m=2$.

\setcounter{equation}{0}
\section{Example: {\mathversion{bold}${\rm RTL}_3^{(+)}(\alpha)$}, 
\\ the three-f\/ield analog of the relativistic Toda lattice}
\label{Sect GRTL example}

In order to provide the reader with an illustration of the bi-Hamiltonian
structure of the system ${\rm RTL}_{m+1}^{(+)}(\alpha)$, we give here  
the corresponding formulas for the case $m=2$, which
is the next in complexity case after the usual relativistic Toda lattice
${\rm RTL}_+(\alpha)$. The equations of motion of the system 
${\rm RTL}_3^{(+)}(\alpha)$ read:
\begin{equation}\label{GRTL3}
\left\{\begin{array}{l}
\dot{d}_k=(1+\alpha d_k)\left(c_k^{(1)}-c_{k-1}^{(1)}\right), 
\vspace{2mm}\\
\dot{c}_k^{(1)}=c_k^{(j)}\left(d_{k+1}-d_k+
\alpha c_{k+1}^{(1)}-\alpha c_{k-1}^{(1)}\right)+
\left(c_k^{(2)}-c_{k-1}^{(2)}\right), 
\vspace{2mm}\\
\dot{c}_k^{(2)}=c_k^{(2)}\left(d_{k+2}-d_k+\alpha c_{k+2}^{(1)}-
\alpha c_{k-1}^{(1)}\right).
\end{array}\right.
\end{equation}
This is, of course, an $O(\alpha)$-perturbation of the system (\ref{GTL3}).
The quadratic Poisson structure of Theorem \ref{r-matrix for GRTL in g+g} 
is charachterized by the following nonvanishing brackets:
\be \label{GRTL3 q br}
\ba{l}
\left\{d_k,c_k^{(1)}\right\}_{2\alpha}  =  -c_k^{(1)}(1+\alpha d_k),  \qquad
\left\{c_k^{(1)},d_{k+1}\right\}_{2\alpha}=-c_k^{(1)}(1+\alpha d_{k+1}), 
\vspace{2mm}\\
\left\{d_k,c_k^{(2)}\right\}_{2\alpha}  =  -c_k^{(2)}(1+\alpha d_k),  \qquad
\left\{c_k^{(2)},d_{k+2}\right\}_{2\alpha}=-c_k^{(2)}(1+\alpha d_{k+2}), 
\vspace{2mm}\\
\left\{c_k^{(1)},c_{k+1}^{(1)}\right\}_{2\alpha}  = 
  -c_k^{(2)}-\alpha c_k^{(1)}c_{k+1}^{(1)},
\qquad 
\left\{c_k^{(1)},c_{k+1}^{(2)}\right\}_{2\alpha}  =  
-\alpha c_k^{(1)}c_{k+1}^{(2)},  
\vspace{2mm}\\
\ds \left\{c_k^{(2)},c_{k+2}^{(1)}\right\}_{2\alpha}=
-\alpha c_k^{(2)}c_{k+2}^{(1)},
\qquad 
\left\{c_k^{(2)},c_{k+1}^{(2)}\right\}_{2\alpha}  = 
-\alpha c_k^{(2)}c_{k+1}^{(2)},
\vspace{2mm}\\
 \left\{c_k^{(2)},c_{k+2}^{(2)}\right\}_{2\alpha}=
-\alpha c_k^{(2)}c_{k+2}^{(2)}.
\ea
\ee
This quadratic bracket is an $O(\alpha)$-perturbation of the {\it linear}
invariant bracket (\ref{GTL3 l br}) of the nonrelativistic system ${\rm TL}_3$. 
The cubic invariant Poisson bracket (see Proposition \ref{GRTL cubic bracket}) 
is characterized by the following nonvanishing brackets:
\[
\left\{d_k,d_{k+1}\right\}_{3\alpha}  =  
 -\alpha c_k^{(1)}\left(1+\alpha d_k\right)
 \left(1+\alpha d_{k+1}\right),
\]
\[
\left\{d_k,c_k^{(1)}\right\}_{3\alpha} = 
 -c_k^{(1)}\left(1+\alpha d_k\right)
 \left(1+\alpha d_k+\alpha^2c_k^{(1)}\right),
\]
\[
\left\{c_k^{(1)},d_{k+1}\right\}_{3\alpha}  = 
 -c_k^{(1)}\left(1+\alpha d_{k+1}\right)
 \left(1+\alpha d_{k+1}+\alpha^2c_k^{(1)}\right),
\]
\[
\left\{d_k,c_{k+1}^{(1)}\right\}_{3\alpha}  = 
 -\alpha\left(c_k^{(2)}+\alpha c_k^{(1)}
 c_{k+1}^{(1)}\right)\left(1+\alpha d_k\right),
\]
\[
\left\{c_k^{(1)},d_{k+2}\right\}_{3\alpha}  = 
 -\alpha\left(c_k^{(2)}+\alpha c_k^{(1)}
 c_{k+1}^{(1)}\right)\left(1+\alpha d_{k+2}\right),
\]
\[
\left\{d_k,c_k^{(2)}\right\}_{3\alpha}  = 
 -c_k^{(2)}\left(1+\alpha d_k\right)
 \left(1+\alpha d_k+\alpha^2c_k^{(1)}\right),
\]
\[
\left\{c_k^{(2)},d_{k+1}\right\}_{3\alpha}  = 
 -\alpha^2c_k^{(2)}
 \left(1+\alpha d_{k+1}\right)\left(c_k^{(1)}-c_{k+1}^{(1)}\right),  
\]
\[
\left\{d_k,c_{k+1}^{(2)}\right\}_{3\alpha}  = 
 -\alpha^2c_k^{(1)}c_{k+1}^{(2)}
 \left(1+\alpha d_k\right),
\]
\[ 
\left\{c_k^{(2)},d_{k+2}\right\}_{3\alpha}  =  
 -c_k^{(2)}\left(1+\alpha d_{k+2}\right)
 \left(1+\alpha d_{k+2}+\alpha^2c_{k+1}^{(1)}\right), 
\]
\[
\left\{c_k^{(2)},d_{k+3}\right\}_{3\alpha}  = 
 -\alpha^2c_k^{(2)}c_{k+2}^{(1)}
 \left(1+\alpha d_{k+3}\right), 
\]
\be
\ba{l}
\ds \left\{c_k^{(1)},c_{k+1}^{(1)}\right\}_{3\alpha} = 
 -\left(1+\alpha d_{k+1}\right)
 \left(c_k^{(2)}+2\alpha c_k^{(1)}c_{k+1}^{(1)}\right)
\vspace{2mm}\\
\ds \qquad -\alpha^2
 \left(c_k^{(1)}+c_{k+1}^{(1)}\right)\left(c_k^{(2)}+\alpha c_k^{(1)}c_{k+1}^{(1)}
 \right),  
\ea
\ee
\[
\left\{c_k^{(1)},c_{k+2}^{(1)}\right\}_{3\alpha}  = 
 -\alpha^2c_k^{(2)}c_{k+2}^{(1)}
 -\alpha^2c_k^{(1)}c_{k+1}^{(2)}-\alpha^3c_k^{(1)}c_{k+1}^{(1)}c_{k+2}^{(1)}, 
\label{GRTL3 c br}
\]
\[ 
\left\{c_k^{(1)},c_k^{(2)}\right\}_{3\alpha}  = 
 -\alpha c_k^{(1)}c_k^{(2)}
 \left(1+\alpha d_{k+1}\right)-\alpha^2c_k^{(2)}
 \left(c_k^{(2)}+\alpha c_k^{(1)}c_{k+1}^{(1)}\right),  
\]
\[
\left\{c_k^{(2)},c_{k+1}^{(1)}\right\}_{3\alpha}  = 
 -\alpha c_k^{(2)}c_{k+1}^{(1)}
 \left(1+\alpha d_{k+1}\right)-\alpha^2c_k^{(2)}
 \left(c_k^{(2)}+\alpha c_k^{(1)}c_{k+1}^{(1)}\right),
\]
\[ 
\left\{c_k^{(1)},c_{k+1}^{(2)}\right\}_{3\alpha}  = 
 -2\alpha c_k^{(1)}c_{k+1}^{(2)}
 \left(1+\alpha d_{k+1}\right)-\alpha^2c_{k+1}^{(2)}\left(c_k^{(2)}
 +\alpha c_k^{(1)}\left(c_k^{(1)}+c_{k+1}^{(1)}\right)\right),
\]
\[
\left\{c_k^{(2)},c_{k+2}^{(1)}\right\}_{3\alpha}  = 
 -2\alpha c_k^{(2)}c_{k+2}^{(1)}
 \left(1+\alpha d_{k+2}\right)-\alpha^2c_k^{(2)}\left(c_{k+1}^{(2)}
 +\alpha c_{k+2}^{(1)}\left(c_{k+1}^{(1)}+c_{k+2}^{(1)}\right)\right),
\]
\[
\left\{c_k^{(1)},c_{k+2}^{(2)}\right\}_{3\alpha}  = 
 -\alpha^2 c_{k+2}^{(2)}
 \left(c_k^{(2)}+\alpha c_k^{(1)}c_{k+1}^{(1)}\right),  
\]
\[
\left\{c_k^{(2)},c_{k+3}^{(1)}\right\}_{3\alpha}  = 
 -\alpha^2 c_k^{(2)} 
 \left(c_{k+2}^{(2)}+\alpha c_{k+2}^{(1)}c_{k+3}^{(1)}\right), 
\]
\[
\left\{c_k^{(2)},c_{k+1}^{(2)}\right\}_{3\alpha}  = 
 -\alpha c_k^{(2)}c_{k+1}^{(2)}
 \left(2+\alpha d_{k+1}+\alpha d_{k+2}+\alpha^2c_k^{(1)}+
 \alpha^2c_{k+2}^{(1)}\right) ,
\]
\[
\left\{c_k^{(2)},c_{k+2}^{(2)}\right\}_{3\alpha}  = 
 -\alpha c_k^{(2)}c_{k+2}^{(2)}
 \left(2+2\alpha d_{k+2}+\alpha^2c_{k+1}^{(1)}+\alpha^2c_{k+2}^{(1)}\right),
\]
\[
\left\{c_k^{(2)},c_{k+3}^{(2)}\right\}_{3\alpha}  = 
 -\alpha^3c_k^{(2)}c_{k+2}^{(1)}c_{k+3}^{(2)} ,
\]
This cubic bracket is again an $O(\alpha)$-perturbation of the {\it linear}
invariant bracket (\ref{GTL3 l br}) of the nonrelativistic system ${\rm TL}_3$. 
However, the linear combination
\[
\alpha^{-1}\left(\{\cdot,\cdot\}_{3\alpha}-\{\cdot,\cdot\}_{2\alpha}\right)
\]
of the both compatible Poisson brackets above leads, in the limit
$\alpha\to 0$, to the {\it quadratic} invariant bracket (\ref{GTL3 q br}) of 
${\rm TL}_3$.

\setcounter{equation}{0}
\section{Conclusions}
\label{Sect conclusions}

The results of this paper conf\/irm once more that the $r$-matrix theory, with 
its various generalizations, provides very convenient means for studying the
Hamiltonian aspects of integrable lattice systems. The relativistic lattice
KP, which seemed for a long time to lie outside of the applicability area
of the $r$-matrix theory, actually f\/its very nicely into this framework. 
Moreover, it allowed us to f\/ind for the f\/irst time the quadratic invariant 
bracket of the relativistic lattice KP, and to establish an amasing fact:
that this attribute remains undeformed by the relativistic deformation.
By the way, this gives also an answer to a problem left open 
in~\cite{suris:SR}, 
concerning the Hamiltonian nature of the ``relativistic Bogoyavlensky
lattices''. Namely, these relativistic systems are Hamiltonian with
respect to the invariant quadratic brackets of their nonrelativistic
counterparts. A detailed account of this latter result will be given 
elsewhere.

\subsection*{Acknowledgement}
I thank B.A.~Kupershmidt for a very stimulating correspondence.

\label{suris-lp}

\end{document}